# Free Space Optical Communication: Challenges and Mitigation Techniques

Hemani Kaushal[1] and Georges Kaddoum[2]
[1]Department of Electrical, Electronics and Communication Engineering, ITM University, Gurgaon, Haryana, India-122017.
[2]Département de génie électrique, École de technologie supérieure, Montréal (Qc), Canada

***Abstract*—In recent years, free space optical (FSO) communication has gained significant importance owing to its unique features: large bandwidth, license free spectrum, high data rate, easy and quick deployability, less power and low mass requirement. FSO communication uses optical carrier in the near infrared (IR) and visible band to establish either terrestrial links within the Earth's atmosphere or inter-satellite/deep space links or ground-to-satellite/satellite-to-ground links. It also find its applications in remote sensing, radio astronomy, military, disaster recovery, last mile access, back-haul for wireless cellular networks and many more. However, despite of great potential of FSO communication, its performance is limited by the adverse effects (viz., absorption, scattering and turbulence) of the atmospheric channel. Out of these three effects, the atmospheric turbulence is a major challenge that may lead to serious degradation in the bit error rate (BER) performance of the system and make the communication link infeasible. This paper presents a comprehensive survey on various challenges faced by FSO communication system for both terrestrial and space links. It will provide details of various performance mitigation techniques in order to have high link availability and reliability of FSO system. The first part of the paper will focus on various types of impairments that poses a serious challenge to the performance of FSO system for both terrestrial and space links. The latter part of the paper will provide the reader with an exhaustive review of various techniques used in FSO system both at physical layer as well as at the upper layers (transport, network or link layer) to combat the adverse effects of the atmosphere. Further, this survey uniquely offers the current literature on FSO coding and modulation schemes using various channel models and detection techniques. It also presents a recently developed technique in FSO system using orbital angular momentum to combat the effect of atmospheric turbulence.**

***Index Terms*—Free space optical communication, atmospheric turbulence, aperture averaging, diversity, adaptive optics, advanced modulation and coding techniques, hybrid RF/FSO, ARQ, routing protocols, orbital angular momentum.**

## I. Introduction

### *A. FSO Communication - An Overview*

In the recent few years, tremendous growth and advancement has been observed in information and communication technologies. With the increase in usage of high speed internet, video-conferencing, live streaming etc., the bandwidth and capacity requirements are increasing drastically. This ever growing demand of increase in data and multimedia services has led to congestion in conventionally used radio frequency (RF) spectrum and arises a need to shift from RF carrier to optical carrier. Unlike RF carrier where spectrum usage is restricted, optical carrier does not require any spectrum licensing and therefore, is an attractive prospect for high bandwidth and capacity applications. "Wireless optical communication" (WOC) is the technology that uses optical carrier to transfer information from one point to another through an unguided channel which may be an atmosphere or free space. WOC is considered as a next frontier for high speed broadband connection as it offers extremely high bandwidth, ease of deployment, unlicensed spectrum allocation, reduced power consumption ($\sim$1/2 of RF), reduced size ($\sim$1/10 the diameter of RF antenna) and improved channel security [1]. It can be classified into two broad categories, namely indoor and outdoor wireless optical communications. Indoor WOC uses IR or visible light for communicating within a building where the possibility of setting up a physical wired connection is cumbersome [2]–[9]. Indoor WOC is classified into four generic system configurations i.e., directed line-of-sight (LOS), non-directed LOS, diffused and tracked. Outdoor WOC is also termed as free space optical (FSO) communication. The FSO communication systems are further classified into terrestrial and space optical links that include building-to-building, ground-to-satellite, satellite-to-ground, satellite-to-satellite, satellite-to-airborne platforms (unmanned aerial vehicles (UAVs) or balloons), [10]–[12] etc. Fig. 1 illustrates the classification of WOC system.

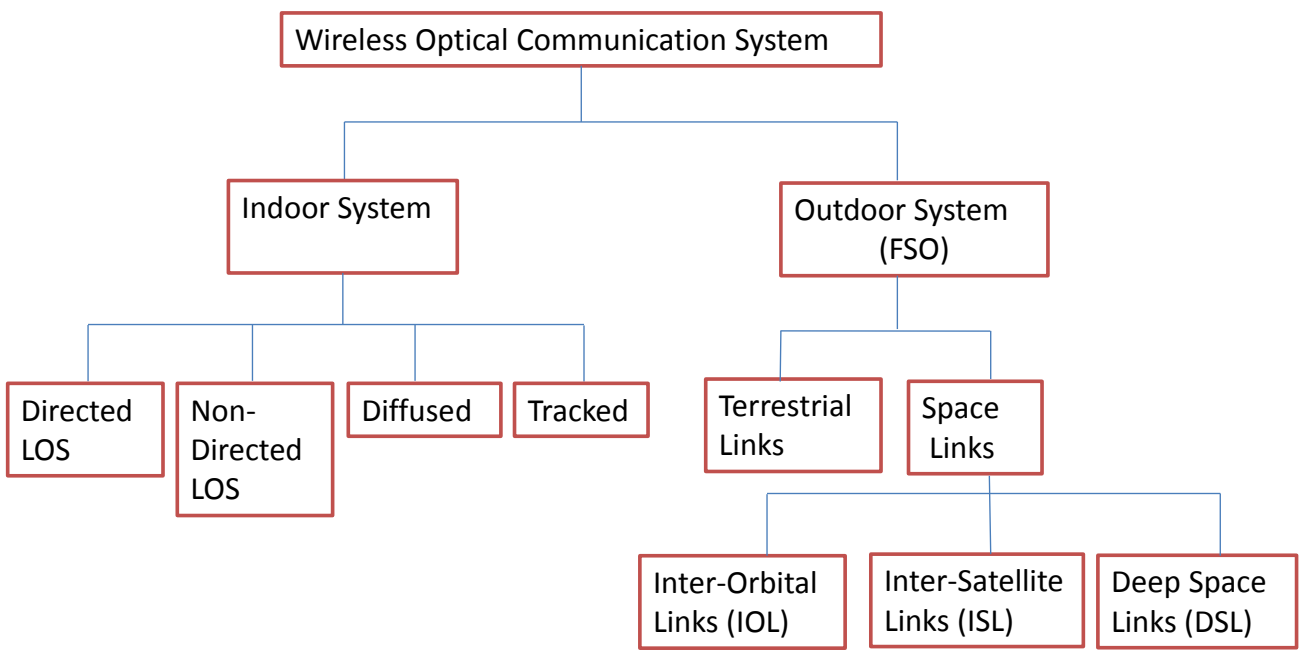

Figure 1. Classification of wireless optical communication system

The FSO communication provides LOS communication owing to its narrow transmit beamwidth and works in visible and IR spectrum. The basic principle of FSO transmission is similar to fiber optic communication except that unlike fiber transmission, in this case the modulated data is transmitted through unguided channel instead of guided optical fiber. The initial work on FSO communication started almost 50 years back for defense and space applications where US military used to send telegraph signal from one point to another using sunlight powered devices. In year 1876, Alexander Graham Bell demonstrated his first wireless telephone system [13], [14] by converting sound waves to electrical telephone signals and transmitted the voice signal over few feets using sunlight as carrier. The device was called "photo-phone" as it was the world's first wireless telephone system. Thereafter, with the discovery of first working laser at Hughes Research Laboratories, Malibu, California in 1960 [15], a great advancement was observed in FSO technology. Large number of experiments were performed in military and aerospace laboratories that demonstrated ground-to-satellite, satellite-to-ground, satellite-to-satellite and ground-to-ground links. It has resulted in various successful experimentations like (i) airborne flight test system (AFTS)- a link between aircraft and ground station at New Mexico [16], (ii) laser cross link system (LCS)- full duplex space-to-space link for geosynchronous system [17], (iii) ground/orbiter lasercom demonstration (GOLD)- first ground-to-space two way communication link [18], [19], (iv) optical communication demonstrator (OCD)- laboratory prototype for demonstrating high speed data transfer from satellite-to-ground, (v) stratospheric optical payload experiment STROPEX (CAPANINA Project)- high bit rate optical downlink from airborne station to transportable optical ground station [20], (vi) Mars laser communications demonstration (MLCD)- provides up to 10 Mbps data transfer between Earth and Mars [21], and (vii) airborne laser optical link (LOLA)- first demonstration of a two-way optical link between high altitude aircraft and GEO satellite (ARTEMIS) [22]. Another mission by NASA is laser communication relay demonstration (LCRD) to be launched in 2017 that will demonstrate optical relay services for near earth and deep space communication missions [23].

Over the last few years, massive expansion in FSO technology has been observed due to huge advancement in opto-electronic components and tremendous growth in the market offering wireless optical devices. FSO communication system seems to be one of the promising technology for addressing the problem of huge bandwidth requirements and "last mile bottleneck". Commercially available FSO equipments provide much higher data rates ranging from 10 Mbps to 10 Gbps [24], [25]. FSO link has been demonstrated in laboratory up to 80 Gbps with average BER of $10^{-6}$ without forward error correction (FEC) coding [26]. Many optical companies like Lightpointe in San Diego, fSONA in Canada, CableFree Wireless Excellence in UK, AirFiber in California etc., provide wide range of wireless optical routers, optical wireless bridges, hybrid wireless bridges, switches etc., that can support enterprise connectivity, last mile access, HDTV broadcast link with almost 100% reliability in adverse weather conditions.

### *B. Advantages of FSO Communication over RF Communication*

FSO communication system offers several advantages over RF system. The major difference between FSO and RF communication arises from the large difference in the wavelength. For FSO system, under clear weather conditions (visibility > 10 miles), the atmospheric transmission window lies in the near infrared wavelength range between 700 nm to 1600 nm. The transmission window for RF system lies between 30 mm to 3 m. Therefore, RF wavelength is thousand of times larger than optical wavelength. This high ratio of wavelength leads to some interesting differences between the two systems as given below:

(I) Huge modulation bandwidth: It is a well known fact that increase in carrier frequency increases the information carrying capacity of a communication system. In RF and microwave communication systems, the allowable bandwidth can be up to 20% of the carrier frequency. In optical communication, even if the bandwidth is taken to be 1% of carrier frequency ($\approx 10^{16}$ Hz), the allowable bandwidth will be 100 THz. This makes the usable bandwidth at an optical frequency in the order of THz which is almost $10^5$ times that of a typical RF carrier [27].

(II) Narrow beam divergence: The beam divergence is proportional to $\lambda/D_R$, where $\lambda$ is the carrier wavelength and $D_R$ the aperture diameter. Thus, the beam spread offered by the optical carrier is narrower than that of RF carrier. This leads to increase in the intensity of signal at the receiver for a given transmitted power. Fig. 2 shows the comparison of beam divergence for optical and RF signals when sent back from Mars towards Earth [28].

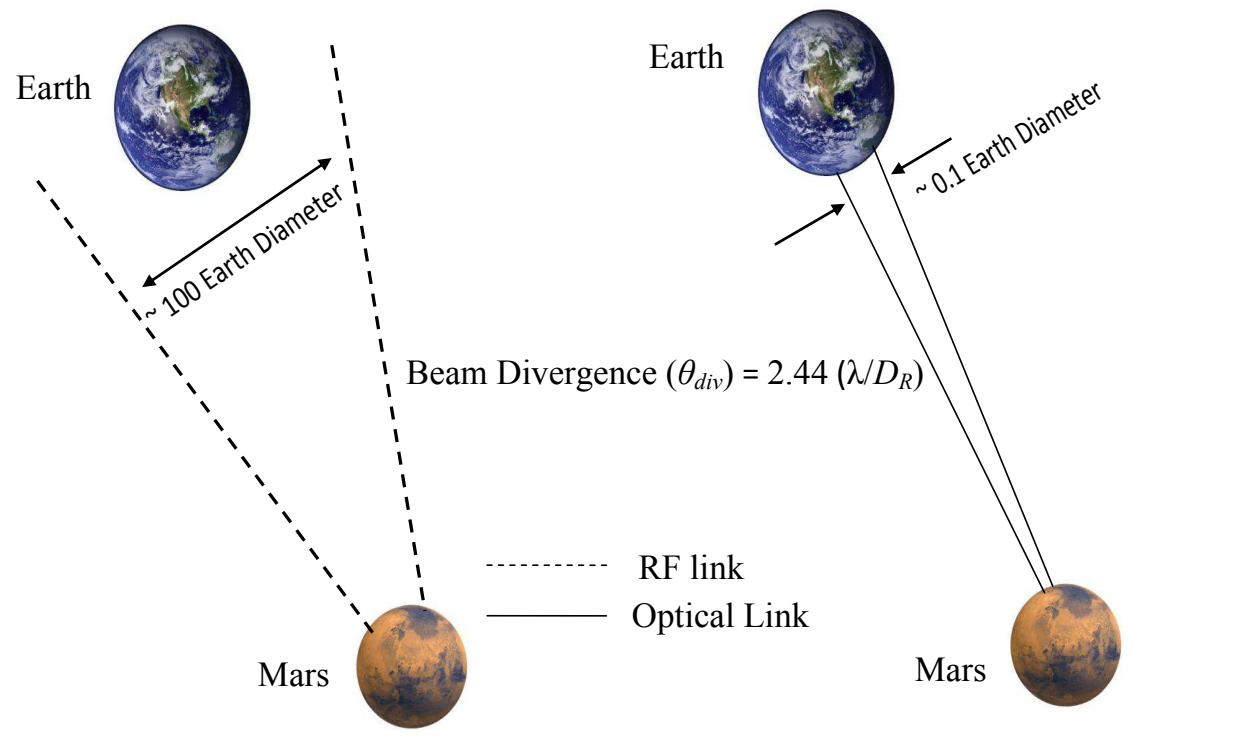

Figure 2. Comparison of optical and RF beam divergence from Mars towards Earth [28]

(III) Less power and mass requirement: For a given transmitter power level, the optical intensity is more at the receiver due to its narrow beam divergence. Thus, a smaller wavelength of optical carrier permits the FSO designer to come up with a system that has smaller antenna than RF system to achieve the same gain (as antenna gain scales inversely proportional to the square of operating wavelength). The typical size for the optical system is 0.3 m vs 1.5 m for the spacecraft antenna [18].

Table I gives the power and mass comparison between optical and RF communication systems using 10 W and 50 W for optical and Ka band systems, respectively at 2.5 Gbps.

| Link | Optical | RF |
|---|---|---|
| **GEO-LEO** | | |
| Antenna Diameter | 10.2 cm (1.0) | 2.2 m (21.6) |
| Mass | 65.3 kg (1.0) | 152.8 kg (2.3) |
| Power | 93.8 W (1.0) | 213.9 W (2.3) |
| **GEO-GEO** | | |
| Antenna Diameter | 13.5 cm (1.0) | 2.1 m (15.6) |
| Mass | 86.4 kg (1.0) | 145.8 kg (1.7) |
| Power | 124.2 W (1.0) | 204.2 W (1.6) |
| **LEO-LEO** | | |
| Antenna Diameter | 3.6 cm (1.0) | 0.8 m (22.2) |
| Mass | 23.0 kg (1.0) | 55.6 kg (2.4) |
| Power | 33.1 W (1.0) | 77.8 W (2.3) |

Table I
COMPARISON OF POWER AND MASS FOR GEOSTATIONARY EARTH ORBIT (GEO) AND LOW EARTH ORBIT (LEO) LINKS USING OPTICAL AND RF COMMUNICATION SYSTEMS (VALUES IN PARENTHESES ARE NORMALIZED TO THE OPTICAL PARAMETERS)

(IV) High directivity: Since the optical wavelength is very small, a very high directivity is obtained with small sized antenna. The directivity of antenna is closely related to its gain. The advantage of optical carrier over RF carrier can be seen from the ratio of antenna directivity as given below

$$\frac{\text{Gain}_{(\text{optical})}}{\text{Gain}_{(\text{RF})}} = \frac{4\pi/\theta^2_{div(\text{optical})}}{4\pi/\theta^2_{div(\text{RF})}}, \tag{1}$$

where $\theta_{div(\text{optical})}$ and $\theta_{div(\text{RF})}$ are the optical and RF beam divergence, respectively and are proportional to $\lambda/D_R$.

(V) Unlicensed spectrum: In RF system, interference from adjacent carrier is the major problem due to spectrum congestion. This requires the need of spectrum licensing by regulatory authorities. But on the other hand, optical system is free from spectrum licensing till now. This reduces the initial set up cost and development time [29].

(VI) High Security: FSO communication can not be detected by spectrum analyzers or RF meters as FSO laser beam is highly directional with very narrow beam divergence. Any kind of interception is therefore very difficult. Unlike RF signal, FSO signal cannot penetrate walls which can therefore prevent eavesdropping [30].

In addition to the above advantages, FSO communication offers secondary benefits as: (i) easily expandable and reduces the size of network segments, (ii) light weight and compact, (iii) easy and quick deployability, and (iv) can be used where fiber optic cables cannot be used. However, despite of many advantages, FSO communication system has its own drawbacks over RF system. The main disadvantage is the requirement of tight acquisition, tracking and pointing (ATP) system due to narrow beam divergence. Also, FSO communication is dependent upon unpredictable atmospheric conditions that can degrade the performance of the system. Another limiting factor, is the position of Sun relative to the laser transmitter and receiver. In a particular alignment, solar background radiations can increase and that will lead to poor system performance [31]. This undoubtedly poses a great challenge to FSO system designers.

### *C. Choice of wavelength in FSO communication*

Wavelength selection in FSO communication is very important design parameter as it affects link performance and detector sensitivity of the system. Since antenna gain is inversely proportion to operating wavelength, therefore, it is more beneficial to operate at lower wavelengths. However, higher wavelengths provide better link quality and lower pointing induced signal fades [32]. Therefore, a careful optimization of operating wavelength in the design of FSO link helps in achieving better performance. The choice of wavelength strongly depends on atmospheric effects, attenuation and background noise power. Further, the availability of transmitter and receiver components, eye safety regulations and cost deeply impacts the selection of wavelength in FSO design process.

The International Commission on Illumination [33] has classified optical radiations into three categories: IR-A (700 nm to 1400 nm), IR-B (1400 nm to 3000 nm) and IR-C (3000 nm to 1 mm). It can sub-classified into (i) near-infrared (NIR) ranging from 750 nm to 1450 nm is a low attenuation window and mainly used for fiber optics, (ii) short-infrared (SIR) ranging from 1400 nm to 3000 nm out of which 1530 nm to 1560 nm is a dominant spectral range for long distance communication, (iii) mid-infrared (MIR) ranging from 3000 nm to 8000 nm is used in military applications for guiding missiles, (iv) long-infrared (LIR) ranging from 8000 nm to 15 $\mu$m is used in thermal imaging, and (v) far-infrared (FIR) is ranging from 15 $\mu$m to 1 mm. Almost all commercially available FSO system are using NIR and SIR wavelength range since these wavelengths are also used in fiber optic communication and their components are readily available in market.

The wavelength selection for FSO communication has to be eye and skin safe as certain wavelengths between 400 nm to 1500 nm can cause potential eye hazards or damage to retina [34]. Under International Electrotechanical Commission (IEC), lasers are classified into four groups from Class 1 to Class 4 depending upon their power and possible hazards [35]. Most of the FSO system uses Class 1 and 1M lasers. For same safety class, FSO system operating at 1500 nm can transmit more than 10 times optical power than system operating at shorter operating wavelengths like 750 nm or 850 nm. It is because cornea, the outer layer of the eye absorb the energy of the light at 1550 nm and does not allow it to focus on retina. Maximum possible exposure (MPE) [36] specifies a certain laser power level up to which person can be exposed without any hazardous effect on eye or skin. Table II summarizes various wavelengths used in practical FSO communication for space applications.

| Mission | Laser | Wavelength | Other parameters | Application |
|---|---|---|---|---|
| Semi-conductor Inter-satellite Link Experiment (SILEX) [37] | AlGaAs laser diode | 830 nm | 60 mW, 25 cm telescope size, 50Mbps, 6 $\mu$rad divergence, direct detection | Inter-satellite communication |
| Ground/Orbiter Lasercomm Demonstration (GOLD) [19] | Argon-ion laser/GaAs laser | Uplink: 514.5 nm Downlink: 830 nm | 13 W, 0.6 m and 1.2m tx. and rx. telescopes size, respectively, 1.024 Mbps, 20 $\mu$rad divergence | Ground-to-satellite link |
| RF Optical System for Aurora (ROSA) [38] | Diode pumped Nd:YVO4 laser | 1064 nm | 6 W, 0.135 m and 10 m tx. and rx. telescopes size, respectively, 320 kbps | Deep space missions |
| Deep Space Optical Link Communications Experiment (DOLCE) [39] | Master oscillator power amplifier (MOPA) | 1058 nm | 1 W, 10-20 Mbps | Inter-satellite/deep space missions |
| Mars Orbiter Laser Altimeter (MOLA) [40] | Diode pumped $Q$ switched Cr:Nd:YAG | 1064 nm | 32.4 W, 420 $\mu$rad divergence, 10 Hz pulse rate, 618 bps, 850 $\mu$rad receiver field of view (FOV) | Altimetry |
| General Atomics Aeronautical Systems (GA-ASI) & TESAT [41] | Nd:YAG | 1064 nm | 2.6 Gbps | Remotely piloted aircraft (RPA) to LEO |
| Altair UAV-to-ground Lasercomm Demonstration [42] | Laser diode | 1550 nm | 200 mW, 2.5 Gbps, 19.5 $\mu$rad jitter error, 10 cm and 1 m uplink and downlink telescopes size, respectively | UAV-to-ground link |
| Mars Polar Lander [43] | AlGaAs laser diode | 880 nm | 400 nJ energy in 100 nsec pulses, 2.5 kHz rate, 128 kbps | Spectroscopy |
| Cloud-Aerosol Lidar and Infrared Pathfinder Satellite Observation (CALIPSO) [44] | Nd:YAG | 532nm/1064nm | 115 mJ energy, 20 Hz rate, 24 ns pulse | Altimetry |
| KIrari's Optical Downlink to Oberpfaffenhofen (KIODO) [45] | AlGaAs laser diode | 847/810 nm | 50 Mbps, 40 cm and 4 m tx. and rx.telescopes size, respectively, 5$\mu$rad divergence | Satellite-to-ground downlink |
| Airborne Laser Optical Link (LOLA) [22] | Lumics fiber laser diode | 800 nm | 300 mW, 50 Mbps | Aircraft and GEO satellite link |
| Tropospheric Emission Spectrometer (TES) [46] | Nd:YAG | 1064 nm | 360 W, 5 cm telescope size, 6.2 Mbps | Interferometry |
| Galileo Optical Experiment (GOPEX) [47] | Nd:YAG | 532 nm | 250 mJ, 12 ns pulse width, 110 $\mu$rad divergence, 0.6 m primary and 0.2 m secondary transmitter telescope size, 12.19 x 12.19 mm CCD array receiver | Deep space missions |
| Engineering Test Satellite VI (ETS-VI) [48] | AlGaAs laser diode (downlink) Argon laser (uplink) | Uplink: 510 nm Downlink: 830 nm | 13.8 mW, 1.024 Mbps bidirectional link, direct detection, 7.5 cm spacecraft telescope size, 1.5 m Earth station telescope | Bi-directional ground-to-satellite link |
| Optical Inter-orbit Communications Engineering Test Satellite (OICETS) [49] | Laser Diode | 819 nm | 200 mW, 2.048 Mbps, direct detection, 25 cm telescope size | Bi-directional Inter-orbit link |
| Solid State Laser Communications in Space (SOLACOS) [50] | Diode pumped Nd:YAG | 1064 nm | 1 W, 650 Mbps return channel and 10 Mbps forward channel, 15 cm telescope size, coherent reception | GEO-GEO link |
| Short Range Optical Inter-satellite Link (SROIL) [51] | Diode pumped Nd:YAG | 1064 nm | 40 W, 1.2 Gbps, 4 cm telescope size, BPSK homodyne detection | Inter-satellite link |
| Mars Laser Communications Demonstration (MLCD) [52] | Fiber laser | 1064 nm and 1076 nm | 5 W, 1- 30 Mbps, 30 cm tx. telescope size and 5 m and 1.6 m rx. telescope size, 64 PPM | Deep space missions |

Table II
WAVELENGTHS USED IN PRACTICAL FSO COMMUNICATION SYSTEMS

### D. Related Surveys and Paper Contributions

Although, FSO communication has been studied in various literatures before, however these surveys still lack to provide the readers with comprehensive detail of every topic. For example, a survey paper by Khalighi and Uysal [53] has elaborated various issues in FSO link according to communication theory prospective. They have presented different types of losses encountered in terrestrial FSO communication, details on FSO transceiver, channel coding, modulation and ways to mitigate fading effects of atmospheric turbulence. However, most of their work is centered around terrestrial FSO communication. Similarly, Bloom et al. [54] have quantitatively covered various aspects for the design of FSO link. This paper has covered primary factors that affect the performance of terrestrial FSO link - atmospheric attenuation, scintillation, alignment or building motion, solar interference and line-of-sight obstructions. It also provide details on transmitter and receiver design, beam propagation models and link budgeting of practical FSO link. Another survey by Demers et al. in [55] solely focused on FSO communication for next generation cellular networks. An introductory paper on terrestrial FSO communication by Ghassemlooy et al. [10] and Henniger et al. [29] provide an overview of various challenges faced in the design of FSO communication. Our survey is also related to various challenges faced in FSO communication system with focus on current status and latest research trends in this field. In our work, we are intending to provide our readers with exhaustive survey of FSO communication for both space and terrestrial links. We have highlighted various performance mitigation techniques at (i) physical level, and (ii) network level.

Our contributions in this paper are listed as follows:

(I) An exhaustive discussion on the basics of FSO communication and its comparison with RF system have been provided to make the readers better understand the switching from conventional RF domain to optical domain. This will make a quick and clear entry point to the topic. It has also provided a comprehensive list of various practical FSO systems deployed for space application till date.

(II) A thorough discussion on various challenges faced in FSO communication system both at terrestrial as well as space links have been presented in this paper. We have also listed out silent features of commonly used atmospheric turbulence profile models.

(III) Effective mitigation techniques for reliable FSO communication have been discussed in this paper. Most of the surveys till date have covered mitigation techniques only at physical layer. This paper uniquely presents atmospheric mitigation techniques both at physical as well as network/transport layer level. A detailed literature survey has been presented for various FSO coding and modulations schemes using various channel models and detection techniques. It has covered the latest survey materials where most of the papers are after the year 2007.

(IV) A recent approach of turbulence mitigation in FSO system using orbital angular momentum (OAM) based on channel coding or adaptive optics has been highlighted.

### E. Paper Organization

The rest of the paper is organized as follows: Section II describes major challenges faced by both terrestrial and space FSO links. Section III presents various techniques both at physical and upper layers (transport, network or link layer) to mitigate the adverse effects of the atmosphere. This section will also give the details of hybrid RF/FSO system that provides a practical solution by backing up the FSO link with low data rate RF link. Section IV gives a recent study on OAM based FSO system using channel coding and adaptive optics for turbulence mitigation. Section V discusses the future scope of FSO technology and finally, the last section will conclude the survey paper.

## II. CHALLENGES IN FSO COMMUNICATION

FSO technology uses atmospheric channel as a propagating medium whose properties are random function of space and time. It makes FSO communication a random phenomena that is dependent on weather and geographical location. Various unpredictable environmental factors like clouds, snow, fog, rain, haze, etc., cause strong attenuation in the optical signal and limit the link distance at which FSO could be deployed. This section will cover various challenges faced by system designer in terrestrial as well as space FSO links.

### A. Terrestrial Links

Terrestrial links include communication between building-to-building, mountain-to-mountain or any other kind of horizontal link between two ground stations. These FSO network can be deployed with point-to-point or point-to-multipoint or ring or mesh topology as shown in Fig. 3. When a laser beam propagates through atmosphere, it experiences power loss due to various factors and a role of system design engineer is to carefully examine the system design requirements in order to combat with the random changes in the atmosphere. For reliable FSO communication, the system design engineer need to have thorough understanding of beam propagation through random atmosphere and its associated losses. The various losses encountered by the optical beam when propagating through the atmospheric optical channel are:

(I) **Absorption and scattering loss:** The loss in the atmospheric channel is mainly due to absorption and scattering process and it is described by Beer's law [56]. At visible and IR wavelengths, the principal atmospheric absorbers are the molecules of water, carbon-dioxide and ozone [57], [58]. The atmospheric absorption is a wavelength dependent phenomenon. Some typical values of molecular absorption coefficients are given in Table III for clear weather conditions. The wavelength range of FSO communication system is chosen to have minimal absorption. This is referred to as atmospheric transmission window. In this window, the attenuation due to molecular or aerosol absorption is less than 0.2 dB/km. There are several transmission windows within the range of 700 - 1600 nm. Majority of FSO systems are designed to operate in the windows of 780 - 850 nm and 1520 - 1600 nm. These wavelengths have been chosen because of the readily availability of the transmitter and detector components at these wavelengths. The wavelength dependance of attenuation under different weather conditions is commonly available in databases like MORTRAN [59], LOWTRAN [60] and HITRAN.

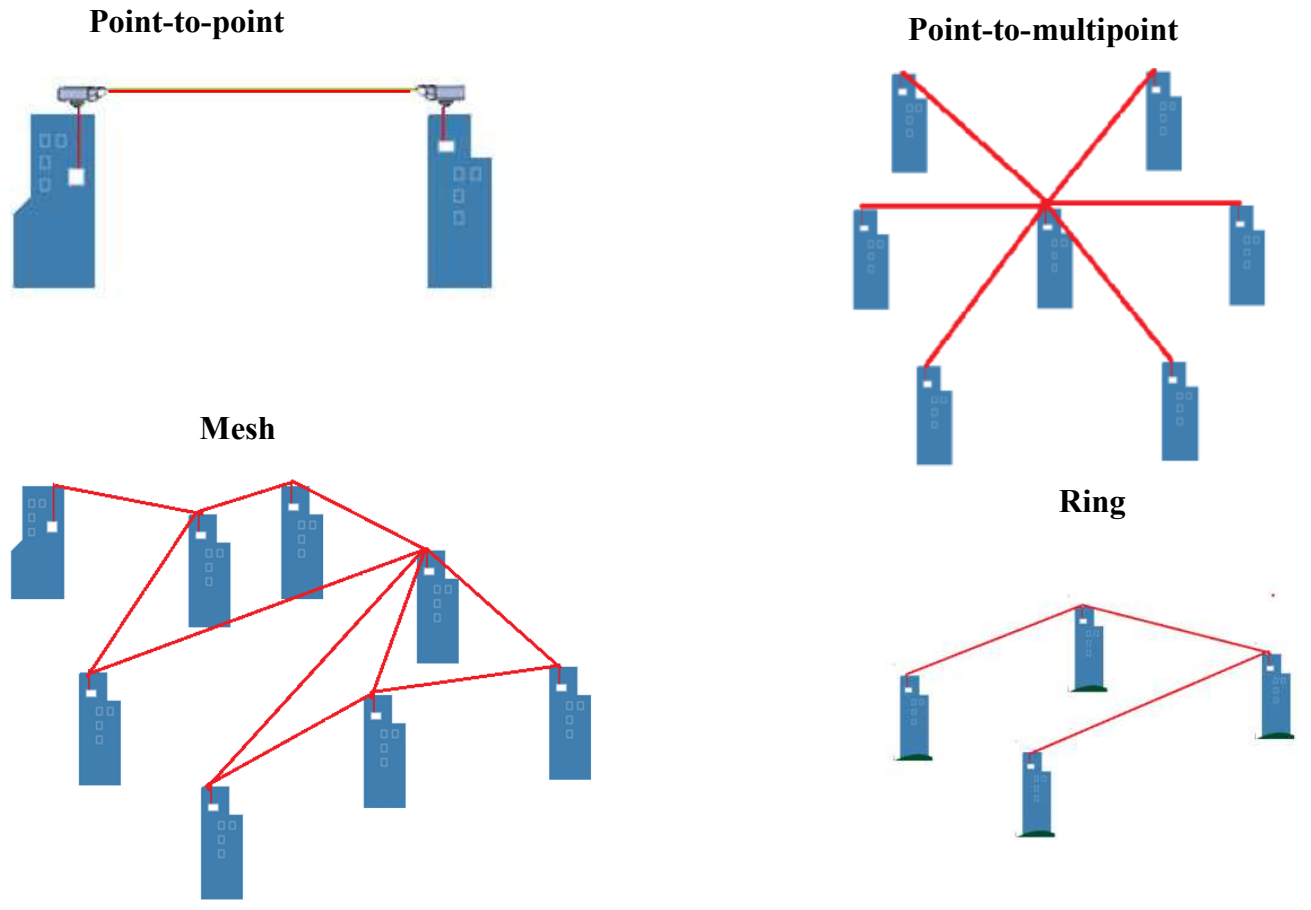

Figure 3. Terrestrial FSO links

| S.No | Wavelength (nm) | Molecular Absorption (dB/km) |
|---|---|---|
| 1. | 550 | 0.13 |
| 2. | 690 | 0.01 |
| 3. | 850 | 0.41 |
| 4. | 1550 | 0.01 |

Table III
MOLECULAR ABSORPTION AT TYPICAL WAVELENGTHS [61]

Scattering of light is also responsible for degrading the performance of FSO system. Like absorption, scattering is also strongly wavelength dependent. If the atmospheric particles are small in comparison with the optical wavelength, then Rayleigh scattering is produced. This scattering is quite prominent for FSO communication around visible or ultraviolet range i.e., wavelengths below 1 $\mu m$. However, it can be neglected at longer wavelengths near IR range. Particles like air molecules and haze cause Rayleigh scattering [62]. If the atmospheric particles size are comparable with the optical wavelength, then Mie scattering is produced. It is dominant near IR wavelength range or longer. Aerosol particles, fog and haze are major contributors of Mie scattering. If the atmospheric particles are much larger than the optical wavelength like in case of rain, snow and hail, then the scattering is better described by geometrical optics model [63], [64].
Total atmospheric attenuation is represented by atmospheric attenuation coefficient which is expressed as combination of absorption and scattering of light. It is therefore expressed as sum of four individual parameters given as

$$\gamma = \alpha_m + \alpha_a + \beta_m + \beta_a, \quad (2)$$

where $\alpha_m$ and $\alpha_a$ are molecular and aerosol absorption coefficients, respectively and $\beta_m$ and $\beta_a$ are molecular and aerosol scattering coefficients, respectively.
Various factors that cause absorption and scattering in FSO system are as follows:

- Fog: The major contributor for atmospheric attenuation is due to fog as it results in both absorption and scattering. During dense fog conditions when the visibility is even less than 50 m, attenuation can be more than 350 dB/km [65]. This clearly shows that it could limit the availability of FSO link. In such cases, very high power lasers with special mitigation techniques help to improve the chances of link availability. Generally, 1550 nm lasers are preferred choice during heavy attenuation because of their high transmitted power. Although, some literature have given evidence that attenuation during dense fog are independent of choice of operating wavelength [66]–[68]. Fig. 4 shows that during heavy fog i.e., low visibility weather condition, all wavelengths (850, 950 and 1550 nm) are almost closely packed following the same pattern implying that specific attenuation is independent of choice of operating wavelength. For light fog i.e., when the visibility range is high (6 km), specific attenuation is quiet less for 1550 nm as compared to 850 nm and 950 nm as shown in Fig. 5. Generally, carrier class availability for all weather and geographical conditions, FSO link distance should be limited to 140-500 meters [69].

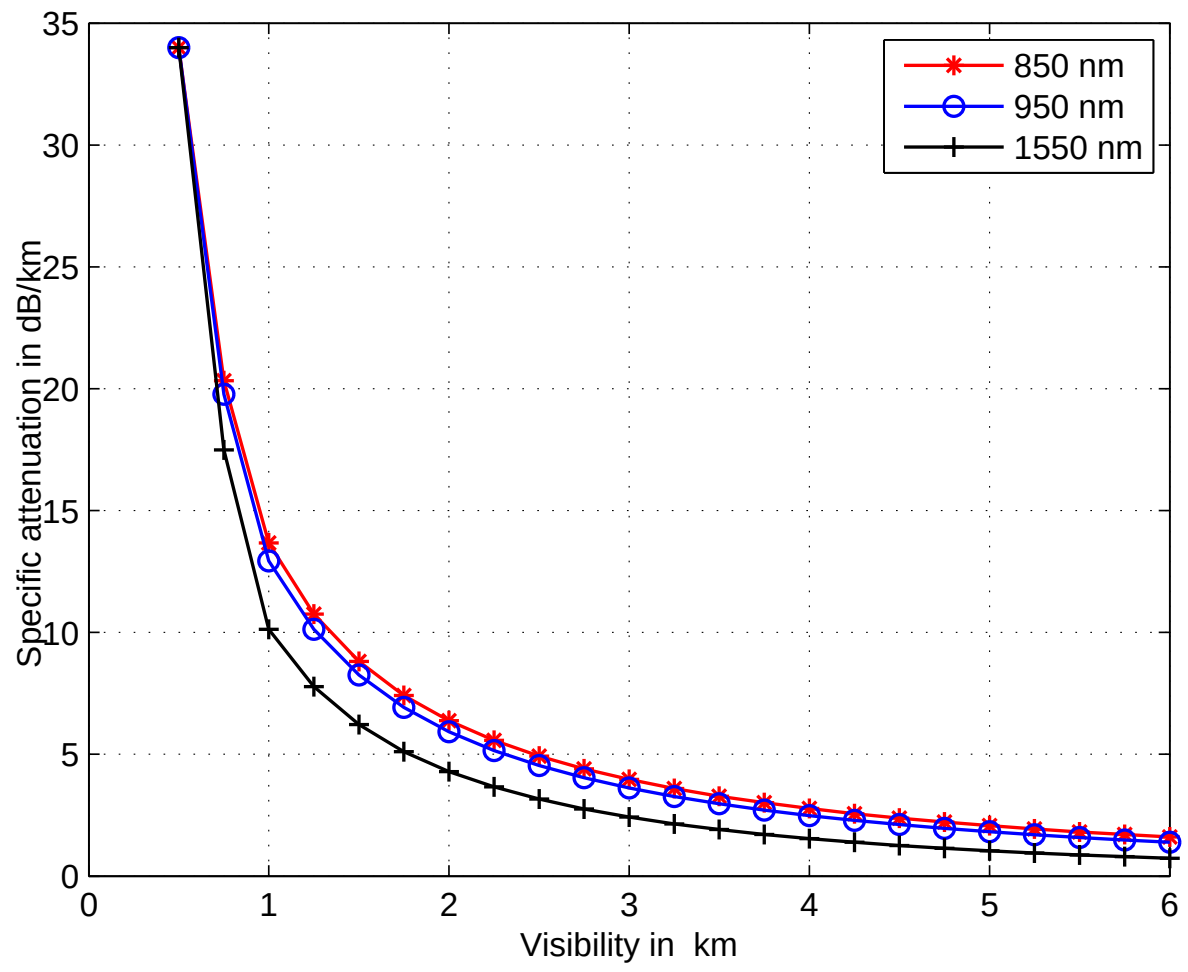

Figure 4. Attenuation vs. visibility during heavy fog

- Rain: The impact of rain is not much pronounced like that fog as rain droplets are significantly larger

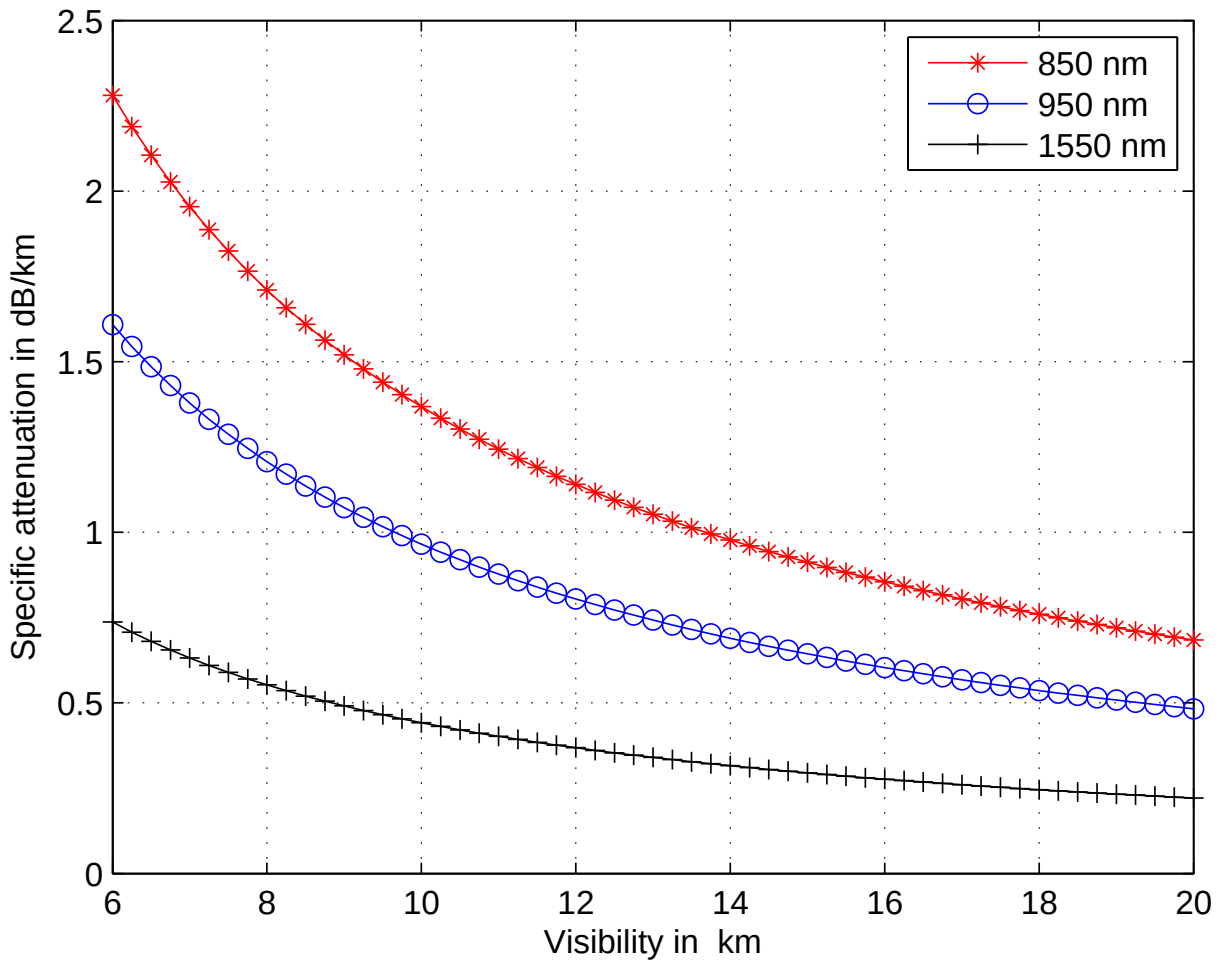

Figure 5. Attenuation vs. visibility during light fog

(100 to 10,000 $\mu m$) in size than the wavelength used in FSO communication. The attenuation loss for light rain (2.5 mm/hr) to heavy rain (25 mm/hr) ranges from 1 dB/km to 10 dB/km for wavelengths around 850 nm and 1500 nm [70], [71]. For this reason, the choice of hybrid RF/FSO system improves the link availability especially for system operating at 10 GHz frequency and above. This topic is discussed in more details in Sec. III.
The modeling of rain attenuation prediction is done using empirical methods proposed by International Telecommunication Union- Radiocommunication sector (ITU-R) for FSO communication [72]. The specific attenuation, $\alpha_{rain}$ (in dB/km) for a FSO link is given by [73]

$$\alpha_{\text{rain}} = k_1 \mathbf{R}^{k_2}, \tag{3}$$

where $\mathbf{R}$ is rain rate in mm/hr and $k_1$and $k_2$ are the model parameters whose values depend upon rain drop size and rain temperature. Table IV gives the values of model parameters as recommended by ITU-R. It is to be noted that rain accompanied by low clouds result in very high attenuation. In order to combat for huge power loss during heavy rain accompanied by low clouds, high power lasers should be used and sufficient link margin greater than 30 dB should be achieved for maximum link availability of FSO system [69].

| Model | Origin | $k_1$ | $k_2$ |
|---|---|---|---|
| Carbonneau | France | 1.076 | 0.67 |
| Japan | Japan | 1.58 | 0.63 |

Table IV
RAIN ATTENUATION MODEL PROPOSED BY ITU-R FOR FSO COMMUNICATION [73].

- Snow: The size of snow particles are between fog and rain particles. Therefore, attenuation due to snow is more than rain but less than fog. During heavy snow, the path of the laser beam is blocked due to increase density of snow flakes in the propagation path or due to the formation of ice on window pane. In this case, its attenuation is comparable to fog ranging between 30-350 dB/km and this can significantly reduce the link availability of FSO system. During heavy snow (150 dB/km), FSO system can given 99.9 % link availability if link margin of approx. 50 dB is chosen [69]. For snow, attenuation is classified into dry and wet snow attenuation. The specific attenuation (dB/km), $\alpha_{snow}$ for snow rate $S$ in mm/hr is given as [72]

$$\alpha_{\text{snow}} = aS^b, \tag{4}$$

where the values of parameters $a$ and $b$ in dry and wet snow are

$$\begin{aligned} \text{Dry snow}: \quad & a = 5.42 \times 10^{-5} + 5.49, \quad b = 1.38, \\ \text{Wet snow}: \quad & a = 1.02 \times 10^{-4} + 3.78, \quad b = 0.72. \end{aligned} \tag{5}$$

(II) **Atmospheric turbulence - horizontal link:** Atmospheric turbulence is a random phenomenon which is caused by variation of temperature and pressure of the atmosphere along the propagation path. It will result in the formation of turbulent cells, also called eddies of different sizes and of different refractive indices. These eddies will act like a prism or lenses and will eventually cause constructive or destructive interference of the propagating beam. It will lead to redistribution of signal energy resulting in random fluctuations in the intensity and phase of the received signal. The intensity fluctuations of the received signal is known as scintillation and is measured in terms of scintillation index (normalized variance of intensity fluctuations), $\sigma_I^2$ given by [74]–[76]

$$\sigma_I^2 = \frac{\langle I^2 \rangle - \langle I \rangle^2}{\langle I \rangle^2} = \frac{\langle I^2 \rangle}{\langle I \rangle^2} - 1, \tag{6}$$

where $I$ is the irradiance (intensity) at some point in the detector plane and the angle bracket $\langle\rangle$ denotes an ensemble average. Scintillation index is expressed as variance of log-amplitude, $\sigma_x^2$ as

$$\sigma_I^2 \approx 4\sigma_x^2, \text{ for } \sigma_x^2 << 1. \tag{7}$$

Scintillation index is function of refractive index structure parameter, $C_n^2$. This parameter determines the strength of turbulence in the atmosphere. Clearly, $C_n^2$ will vary with time of day, geographical location and height. For near ground horizontal link, value of $C_n^2$ is almost constant and its typical value in case of weak turbulence is $10^{-17}$ $\text{m}^{-3/2}$ and for strong turbulence it can be up to $10^{-13}$ $\text{m}^{-3/2}$ or greater. Various empirical models of $C_n^2$ have been proposed to estimate the turbulence profile that are based on experimental measurements carried out at variety of geographical locations, time of day, wind speed, terrain type, etc [77]. Some of the commonly used models are presented in Table V.

| Models | Range | Comments |
|---|---|---|
| PAMELA Model [78] | Long (few tens of kms) | - Robust model for different terrains and weather type<br>- Sensitive to wind speed<br>- Does not perform well over marine/overseas environment |
| NSLOT Model [79] | Long (few tens of kms) | - More accurate model for marine propagation<br>- Surface roughness is 'hard-wired' in this model<br>-Temperature inversion i.e., $(T_{air} - T_{sur} > 0)$ is problematic |
| Fried Model [80] | Short (in meters) | - Support weak, strong and moderate turbulence |
| Hufnagel and Stanley Model [81] | Long (few tens of kms) | - $C_n^2$ is proportional to $h^{-1}$<br>- Not suitable for various site conditions |
| Hufnagel Valley Model [82]–[84] | Long (few tens of kms) | - Most popular model as it allows easy variation of daytime and night time profile by varying various site parameters like wind speed, iso-planatic angle and altitude<br>- Best suited for ground-to-satellite uplink<br>- HV 5/7 is a generally used to describe $C_n^2$ profile during day time. HV5/7 yields a coherence length of 5 cm and isoplanatic angle of 7$\mu$rad at 0.5 $\mu$m wavelength |
| Gurvich Model [85] | Long (few tens of kms) | - Covers all regimes of turbulence from weak, moderate to strong<br>- $C_n^2$ dependance on altitude, *h*, follows power law i.e., $C_n^2 \propto h^{-n}$ where *n* could be 4/3, 2/3 or 0 for unstable, neutral or stable atmospheric conditions, respectively. |
| Von Karman-Tatarski Model [86], [87] | Medium (few kms) | - Make use of phase peturbations of laser beam to estimate inner and outer scale of turbulence<br>- Sensitive to change in temperature difference |
| Greenwood Model [88], [89] | Long (few tens of kms) | - Night time turbulence model for astronomical imaging from mountaintop sites |
| Submarine Laser Communication (SLC) [90] Model | Long (few tens of kms) | - Well suited for day time turbulence profile at inland sites<br>- Developed for AMOS observatory in Maui, Hawaii |
| Clear 1 [91] | Long (few tens of kms) | - Well suited for night time turbulent profile<br>- Averages and statistically interpolate radiosonde observation measurements obtained from large number of meteorological conditions |
| Aeronomy Laboratory Model (ALM) [92] | Long (few tens of kms) | - Shows good agreement with radar measurements<br>- Based on relationship proposed by Tatarski [87] and works well with radiosonde data |
| AFRL Radiosonde Model [93] | Long (few tens of kms) | - Similar to ALM but with simpler construction and more accurate results as two seperate models are used for troposphere and stratosphere<br>- Daytime measurements could give erroneous results due to solar heating of thermosonde probes |

Table V
TURBULENCE PROFILE MODELS FOR $C_n^2$

For weak turbulence i.e., $\sigma_I^2 < 1$, the intensity statistics is given by log-normal distribution. For strong turbulence, $\sigma_I^2 \geq 1$, the field amplitude is Rayleigh distributed which means negative exponential statistics for the intensity [80]. Besides these two models, a number of other statistical models [94] are used in literature to describe the scintillation statistics in either a regime of strong turbulence ($K$ model) or all the regimes ($I$-$K$ and Gamma-Gamma [95] models). For 3 $<\sigma_I^2 <$ 4, the intensity statistics is given by $K$ distribution. Another generalized form of $K$ distribution that is applicable to all conditions of atmospheric turbulence is *I-K* distribution. However, *I-K* distribution is difficult to express in closed form expressions. In that case, the Gamma-Gamma distribution is used to successfully describe the scintillation statistics for weak to strong turbulence [96]. Although Gamma-Gamma distribution is most widely used to study the performance of FSO system, however, in recent work proposed by Chatzidiamantis et al. [97], DoubleWeibull distribution is suggested to be more accurate model for atmospheric turbulence than the Gamma-Gamma distribution, particularly for the cases of moderate and strong turbulence. Another very latest turbulence model proposed in [98] is Double Generalized Gamma (Double GG) distribution which is suitable for all regimes of turbulence and it covers almost all the existing statistical models of irradiance fluctuations as special cases.

Scintillation index for weak turbulence in case of plane and spherical waves is expressed as

$$\sigma_I^2 = \sigma_R^2 = 1.23 C_n^2 k^{7/6} L^{11/6} \text{ for plane wave,} \quad (8)$$

$$\sigma_I^2 = 0.4\sigma_R^2 = 0.5 C_n^2 k^{7/6} L^{11/6} \text{ for spherical wave,} \quad (9)$$

where $k$ is wave number ($2\pi/\lambda$), $\sigma_R^2$ the Rytov variance, $L$ the link distance. It is clear from Eqs.(8) and (9) that for given link distance in case of weak turbulence conditions, irradiance fluctuations will decrease at longer wavelength. Scintillation index in case of strong turbulence is given by

$$\sigma_I^2 = 1 + \frac{0.86}{\sigma_R^{4/5}}, \ \sigma_R^2 >> 1 \text{ for plane wave,} \quad (10)$$

$$\sigma_I^2 = 1 + \frac{2.73}{\sigma_R^{4/5}}, \ \sigma_R^2 >> 1 \text{ for spherical wave.} \quad (11)$$

Depending on the size of turbulent eddies and transmitter beam size, three types of atmospheric turbulence effects can be identified:

- Turbulence induced beam wander: Beam wander is a phenomenon which is experienced *when the size of turbulence eddies are larger than the beam size*. It will result in random deflection of the beam from its propagating path and leads to link failure. The rms beam wander displacement is function of link length ($L$), operating wavelength ($\lambda$), and initial beam size ($W_o$) and is given by $\sigma_{BW}^2 = 1.44 C_n^2 L^2 W_o^{-1/3}$ [99].

- Turbulence induced beam spreading: Beam spreading takes into account *when the size of the eddies are smaller than the beam size*. In this case, the incoming beam will be diffracted and scattered independently leading to distortion of the received wavefront.
- Turbulence induced beam scintillation: *When the eddy size is of the order of beam size*, then the eddies will act like lens that will focus and de-focus the incoming beam. This will result in temporal and spatial irradiance fluctuations of the laser beam and is major cause of degradation in the performance of FSO system.

(III) **Beam divergence loss:** As the optical beam propagates through the atmosphere, beam divergence is caused by diffraction near the receiver aperture. Some fraction of the transmitted beam will not be collected by the receiver and it will result in beam divergence loss/geometrical loss. This loss increases with the link length unless the size of the receiver collection aperture is increased or receiver diversity is employed.

(IV) **Ambient light:** Light from luminous bodies like Sun, moon or fluorescent objects produces shot noise and it interferes with the noise present at the detector. Both the detector noise and ambient noise contribute to total receiver noise and this result in some noise flicker or disturbances. In order to minimize noise from ambient light sources, FSO system should be operated at higher wavelengths. The solar irradiance spectrum ranges from 300 nm to more than 2000 nm showing peak at around 500 nm. Thereafter, the pattern decreases with the increase in the wavelength [100].

(V) **Mis-alignment or building sway:** The optical beam used in FSO communication is highly directional with very narrow beam divergence. Also, receivers used in FSO links have limited FOV. Therefore, in order to have 100% availability of FSO link, it is very essential to maintain a constant LOS connection between the transmitter and receiver. However, the building over which FSO transceivers are mounted are in constant motion due to variety of factors such as thermal expansion, vibrations or high wind velocity. This will lead to failure of FSO link due to mis-alignment or building sway. It poses a great challenge for transceiver alignment and one needs to have very accurate pointing and tracking mechanisms to overcome FSO link failure.

Besides the above mentioned factors, there could be other reasons for link failure. Since FSO system requires LOS communication, any kind of physical obstruction can block the beam path and cause a short and temporary interruptions of the received signal. This adverse effect can be taken care of by proper choice of system design parameters like beam divergence, transmitter power, operating wavelength, transmitter and receiver FOV [101].

### B. Space Links

Space links include both ground-to-satellite/ satellite-to-ground links, inter-satellite links and deep space links. Links between LEO to GEO are used for transmitting gathered data from LEO to GEO which in turn transmits data to other part of the Earth as shown in Fig. 6.

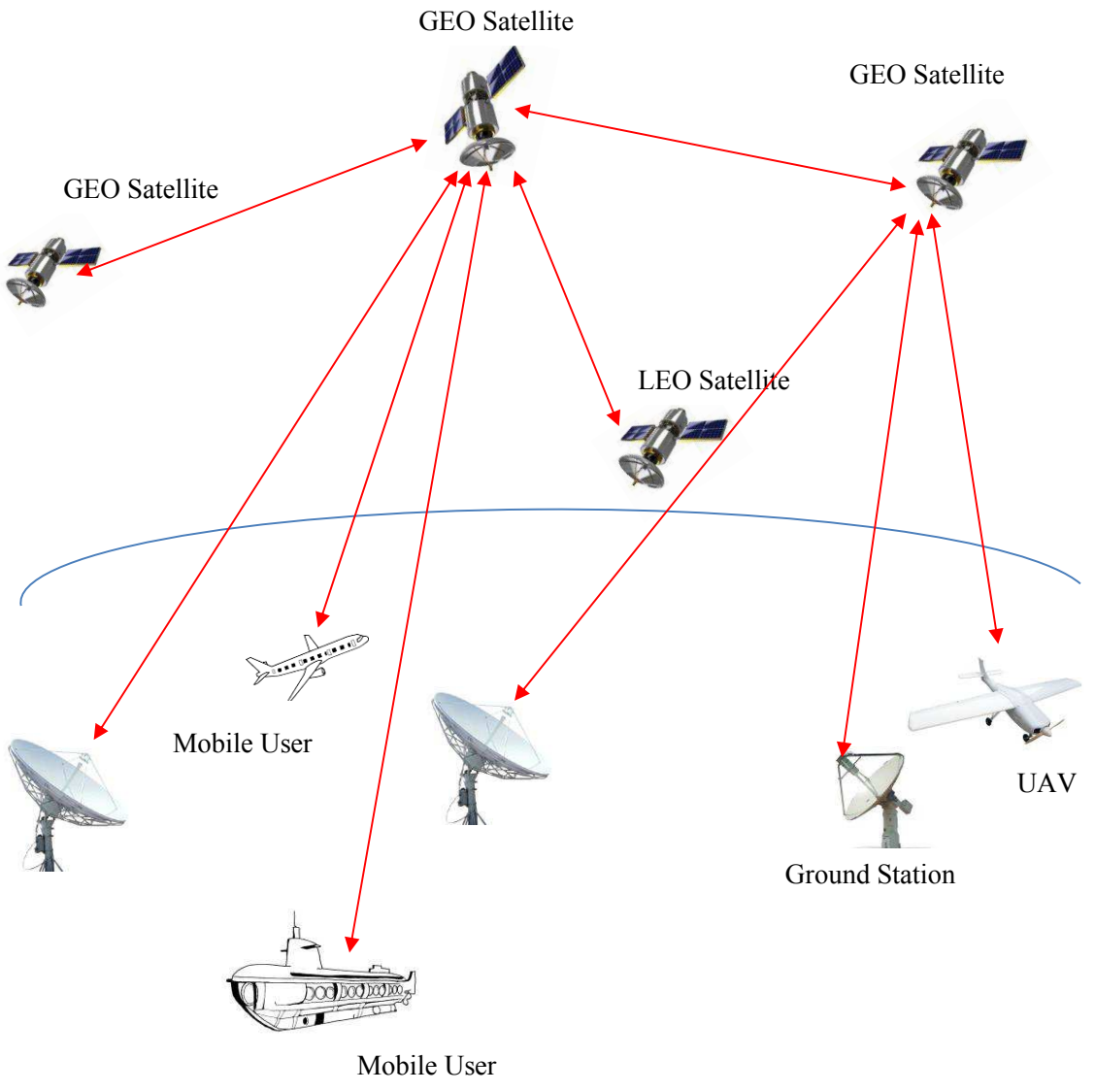

Figure 6. Space FSO links

Many researchers in US, Europe and Japan are investigating space-to-ground links using LEO (mobile FSO link). Optical Inter-Orbit Communications Engineering Test Satellite (OICETS) was the first successful bi-directional optical link between KIRARI, the Japanese satellite and ESA's Artemis in 2001 [102]. Also, successful operational inter-satellite optical link was established between Artemis and Spot4 via SILEX system [103]. An optical link between two LEO orbiting satellites, Terra SAR-X and NFIRE, at 5.5 Gbps on a total distance of 5500 km and at a speed of 25, 000 km/hr has been established in 2008. A 2.5 Gbps experiment was performed successfully between LEO satellite and ground station at 1 W laser power, 1064 nm wavelength using BPSK modulation scheme [104]. An optical link at 2.5 Gbps was demonstrated by NASA between ground station and UAV achieving a BER of $10^{-9}$ at 1550 nm wavelength [42]. These space links have to face severe challenges due to adverse atmospheric effects (in case of ground-to-satellite/satellite-to-ground links) as discussed in previous section as well as require very tight acquisition, tracking and pointing owing to its narrow beam width.

(I) **Pointing loss:** Pointing error is one of the major challenge in FSO communication that can result in link failure. It is very essential to maintain pointing and acquisition throughout the duration of communication. It could arise due to many reasons such as satellite vibration or platform jitter or any kind of stress in electronic or mechanical devices. The effect of satellite vibration in FSO system is described in [105]–[108]. Pointing error can also be caused due to atmospheric turbulence induced beam wander effect which can displace the beam from its transmit path [109]. In any of the case, pointing error will increase the chances of link failure or can significantly reduce the amount of received power at the receiver resulting in high probability of error. In order to achieve sub micro-radian pointing accuracy, proper care has to be taken to make the assembly vibration free and maintain sufficient bandwidth control and dynamic range in order to compensate for residual jitter [110].
Total pointing error, $\sigma_p$ is sum of tracking error, $\sigma_{track}$ and point ahead error, $\sigma_{pa}$ i.e., $\sigma = \sigma_{track} + \sigma_{pa}$. Tracking error is primarily due to the noise associated with tracking sensors or due to disturbances arising from mechanical vibration of satellite. Point ahead error occurs if the calculation of point ahead angle did not allow sufficient transit time from satellite-to-ground and back again. It could be due to error in Ephemeris data or point ahead sensor error or calibration error or waveform deformation. Pointing error loss is more when tracking LEO than GEO satellite [111], [112]. Also, loss due pointing error is more significant at visible wavelength and decreases at higher wavelength due to inherent broadening of beam. Pointing error has significant impact on BER performance of FSO system. Fig. 7 shows the BER performance in the presence of random jitter.

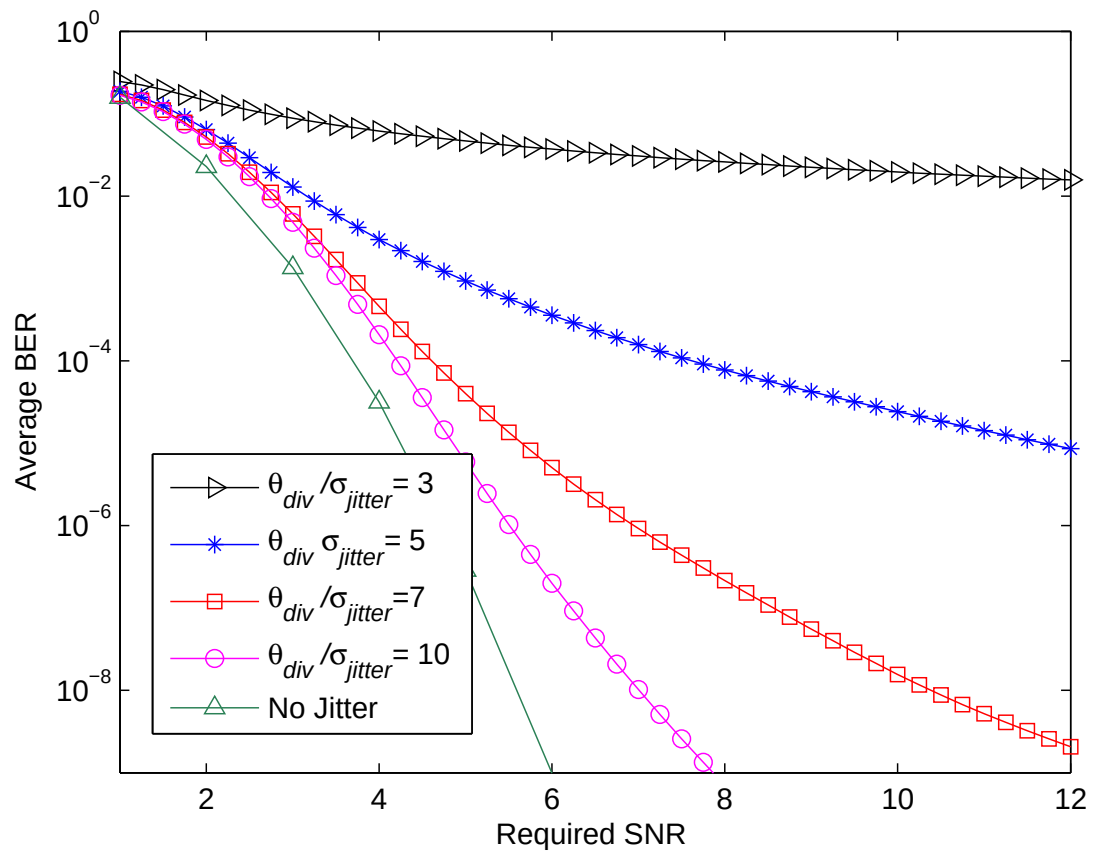

Figure 7. BER vs SNR for different values of ratio of beam divergence angle to random jitter

(II) **Atmospheric turbulence- vertical link:** For vertical links, the value of $C_n^2$ changes with altitude $h$ unlike horizontal link where its value is assumed to be constant. With the increase in the altitude, the value of $C_n^2$ decreases at the rate of $h^{-4/3}$. Therefore, for vertical links, the value of $C_n^2$ has to integrated over the complete propagation path extending from height of the receiver above sea level to the top of the atmosphere (roughly up to 40 kms). Due to this reason, the effect of atmospheric turbulence from ground-to-satellite (uplink) is different from satellite-to-ground (downlink).
Various empirical models of $C_n^2$ have been proposed in [113], [114] that describe the strength of the atmospheric turbulence with respect to the altitude (as mentioned in Table V). The most widely used model for vertical link is Hufnagel Valley Boundary (HVB) model [115] given by

$$\begin{aligned} C_n^2(h) &= 0.00594\Big[\left(\tfrac{V}{27}\right)^2\left(10^{-5}h\right)^{10}\exp\left(-h/1000\right) \\ &\quad +2.7\times10^{-16}\exp\left(-\tfrac{h}{1500}\right) \\ &\quad +A\exp\left(-\tfrac{h}{100}\right)\Big]\ m^{-2/3}, \end{aligned} \quad (12)$$

where $V^2$ is the mean square value of the wind speed in m/s, $h$ is the altitude in meters and $A$ is a parameter whose value can be adjusted to fit various site conditions. The parameter $A$ is given as

$$\begin{aligned} A &= 1.29\times10^{-12}r_0^{-5/3}\lambda^2 - 1.61\times10^{-13}\theta_0^{-5/3}\lambda^2 \\ &\quad +3.89\times10^{-15}. \end{aligned} \quad (13)$$

In the above equation, $\theta_0$ is the isoplanatic angle [84] (angular distance over which the atmospheric turbulence is essentially unchanged) and $r_0$ is the atmospheric coherence length [116]. The coherence length of the atmosphere is an important parameter that is dependent upon operating wavelength, $C_n^2$ and zenith angle $\theta$. For plane wave propagating from altitude $h_o$ to $(h_o + L)$ (downlink), it is given as

$$r_0 = \left[0.423k^2\sec(\theta)\int_{h_0}^{h_o+L} C_n^2(h)\,dh\right]^{-3/5}. \quad (14)$$

For spherical wave (downlink), it is expressed as

$$r_0 = \left[0.423k^2\sec(\theta)\int_{h_0}^{h_o+L} C_n^2(h)\left\{\frac{L+h_o-h}{L}\right\}^{5/3}dh\right]^{-3/5}. \quad (15)$$

It is clear from above expressions that $r_0$ varies as $\lambda^{6/5}$, therefore, FSO link operating at higher wavelengths will have less impact of turbulence than at lower wavelengths. For the uplink, if transmitter beam size $W_0$ is of the order of $r_0$, significant beam wander takes place. For downlink, angle of arrival fluctuation [74] increases as the value of $r_0$ decreases. In case of weak turbulence, scintillation index for plane wave (downlink) can be written in terms of refractive index structure parameter, $C_n^2$ as

$$\sigma_I^2 = \sigma_R^2 \approx 2.24k^{7/6}\left(\sec(\theta)\right)^{11/6}\int_{h_0}^{h_o+L} C_n^2(h)\,h^{5/6}dh. \quad (16)$$

It should be noted that weak fluctuation theory does not hold true for larger zenith angles and smaller wavelengths. In that case, scintillation index for moderate to strong turbulence holds well and is given by [117]

$$\sigma_I^2 = \exp\left[\frac{0.49\sigma_R^2}{\left(1+1.11\sigma_R^{12/5}\right)^{7/6}} + \frac{0.51\sigma_R^2}{\left(1+0.69\sigma_R^{12/5}\right)^{7/6}}\right] - 1. \quad (17)$$

(III) **Background noise:** The main sources of background noise are: (a) diffused extended background noise from the atmosphere, (b) background noise from the Sun and other stellar (point) objects and (c) scattered light collected by the receiver [118]. The background noise can be controlled by limiting the receiver optical bandwidth. Single optical filter with very narrow bandwidth in the order of approx. 0.05 nm can be used to control the amount of background noise. In addition, the other sources of noise in FSO system are detector dark current, signal shot noise and thermal noise. Total noise contribution is sum of background noise and noise due to other sources.

(IV) **Atmospheric seeing:** The perturbations of the optical beam associated with coherence length of the atmosphere, $r_0$ is referred as atmospheric seeing effect. When $r_0$ is significantly smaller than the receiver aperture diameter $D_R$, then it leads to the blurring of received signal which is known as astronomical seeing which is given as $\lambda/r_0$ [119]. For a perfect optical collection system, the spot size of the received signal in the focal plane of the receiver is expressed as $(2.44F\lambda/D_R)$ where $F$ is the focal length of receiver collecting optics. When the optical beam propagates through atmosphere, then $D_R$ is replaced by $r_0$ and therefore, the related signal spot size at the focal plane is increased by the ratio $D_R/r_0$ which effectively leads to increase in the background noise. Also, larger FOV at the receiver can limit the electrical bandwidth of the receiver thereby limiting the data rate. This problem can be taken care of by use of adaptive optics or array detectors.

(V) **Angle of arrival fluctuations:** Due to the presence of turbulence in the atmosphere, the laser beam wavefront arriving at the receiver will be distorted. This will lead to spot motion or image dancing at the focal plane of the receiver. This effect is called angle of arrival fluctuations. However, this effect can be compensated by use of adaptive optics or fast beam steering mirror. For plane wave, the variance of angle of arrival fluctuations, $\langle\beta^2\rangle$ is expressed as [75]

$$\langle\beta^2\rangle = \begin{cases} 1.64C_n^2 L l_o^{-1/3}, & D_R \ll l_o \\ 2.91C_n^2 L D_R^{-1/3}, & D_R \gg l_o \end{cases}, \tag{18}$$

where $D_R$ is the diameter of collecting lens and $l_o$ the inner scale of turbulent eddy.

## III. Mitigation Techniques

Atmospheric channel causes degradation in the quality of received signal which deteriorate the BER performance of the FSO system. In order to improve the reliability of FSO system for all weather conditions, various types of mitigation techniques are employed. Mitigation technique can be used either at physical layer or at network layer. Multiple beam transmission, increasing receiver FOV, adaptive optics, relay transmission, hybrid RF/FSO etc., are some of the mitigation techniques used at physical layer. Packet re-transmission (in FSO link or network), network re-routing, quality of service (QoS) control, data re-play are some of the methods used in network layer in order to improve the performance and availability of FSO system. Fig. 8 gives various mitigation techniques used in FSO communication.

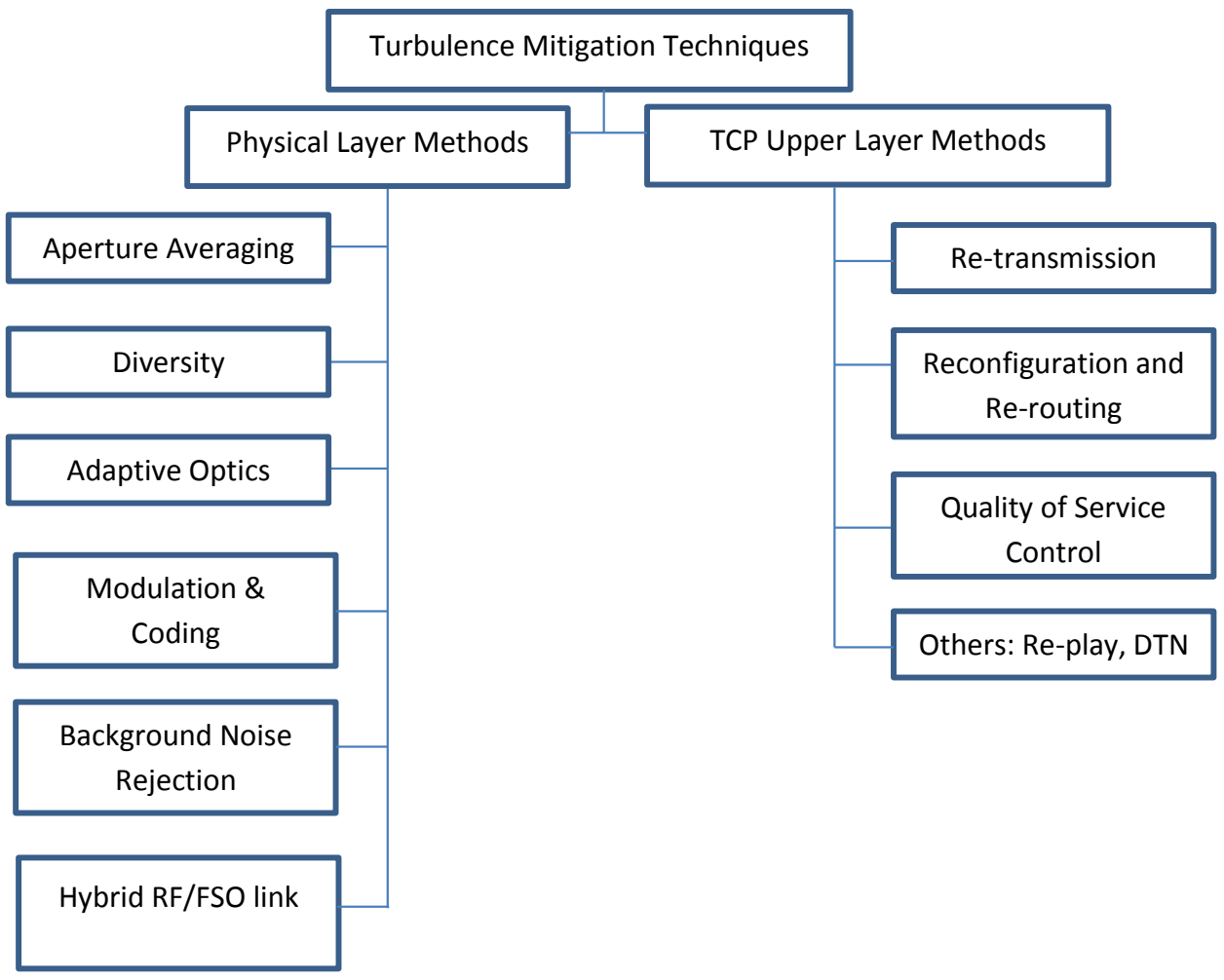

Figure 8. Various techniques for mitigating atmospheric turbulence

### A. Physical Layer Methods

- **Aperture Averaging:** This technique is used to mitigate the effect of atmospheric turbulence by increasing the size of the receiver aperture that average out relatively fast fluctuations caused by the small-size eddies and helps in reducing channel fading. The parameter that quantify reduction in fading due to aperture averaging is called aperture averaging factor, $A$. The parameter $A$ is defined as ratio of variance of the signal fluctuations from a receiver with aperture diameter $D_R$ to that from a receiver with an infinite small aperture i.e.,

$$A = \frac{\sigma_I^2(D_R)}{\sigma_I^2(0)}. \tag{19}$$

Various approximations for the aperture averaging factor have been given by Churnside [120], Andrew [121], etc. The Churnside approximation of aperture averaging factor for plane wave in weak turbulence region is given by

$$A = \left[1 + 1.07\left(\frac{kD_R^2}{4L}\right)^{7/6}\right]^{-1}. \tag{20}$$

Similarly, the approximated value for the aperture averaging factor for spherical wave is given:

$$A = \left[1 + 0.214\left(\frac{kD_R^2}{4L}\right)^{7/6}\right]^{-1}. \tag{21}$$

Therefore, increasing the aperture diameter reduces atmospheric scintillation and improves the BER performance of the system. Various literature on aperture averaging can be found in [122]–[126]. NASA conducted an experiment on aperture averaging using 1550 nm operating wavelength through transmit aperture of 2.5 cm for propagation path of 1 km using Gaussian beam. The receiver aperture diameter was varied up to the size of 8 inches. It was seen from the results that BER was around $10^{-3}$ for small aperture diameter of 2 inches. However, with increase in aperture diameter to 8 inches, BER performance reached up to $10^{-12}$ or even better [127]. In [128], irradiance statistics of a Gaussian beam propagating through turbulent atmosphere along a horizontal path was investigated. It was observed that for moderate to strong turbulence regime, Gamma-Gamma distribution provides the best fit to irradiance statistics. In case of aperture diameter larger than the coherence length of the atmosphere, the irradiance statistics appear to be log-normal.

It should be noted that increase in the receiver aperture area will also increase the amount of background noise collected by the receiver. Therefore, an optimum choice of aperture diameter has to be made in order to increase the power efficiency in FSO system.

- **Diversity:** Diversity technique for mitigating the effect of turbulence in the atmosphere can operate on time, frequency and space. In this case, instead of single large aperture, an array of smaller receiver aperture is used so that multiple copies of the signal that are mutually uncorrelated can be transmitted either in time or frequency or space. This will improve the link availability and BER performance of the system. It also limits the need of active tracking due to laser misalignment [120], [129], [130]. GOLD demonstration in 1998 showed that using four 514.5 nm multiple beams for uplink transmission, scintillation index was drastically improved. It was reported that the value of scintillation was 0.12 with two beams, however, its value reduces to 0.045 with four beams [131]. In order to achieve the full advantage due to spatial diversity, the antennas separation at transmitter or receiver should be at least or greater than coherence length of the atmosphere to make the multiple beams independent or at least uncorrelated. The effect of correlation between multiple beams is presented in [129] where it shows that a correlation of 0.25 among three transmit apertures can decrease the diversity order by 1. The situation worsens with the higher order correlation. This is in contrast to RF communication where only full spatial correlation results in loss of diversity. Also, the gain due to diversity is more pronounced at high turbulence level than at lower values [132], [133].

In case of receiver diversity (SIMO- single input multiple output), diversity gain is achieved by averaging over multiple independent signal paths. The signals can be combined at the receiver using selection combining (SC) or equal gain combining (EGC) or maximal ratio combining (MRC). SC is simpler as compared to other two, but gain in this case is low. The gain achieved through MRC is slightly higher than EGC, but at the expense of complexity and cost. Therefore, implementation of EGC is preferred over MRC due to its simplicity and comparable performance. In case of receiver diversity using intensity modulation/direct detection (IM/DD), it has been verified using wave-optics and Monte Carlo simulations that effect of correlation corresponding to small scale turbulence can be neglected, irrespective of atmospheric turbulence condition [134]. For transmit diversity (MISO- multiple input single output), special space time codes such as optical Alamouti code is used [96], [135]. This code is designed for only two transmit antenna but can be extended to more number of antennas. The performance of optical MIMO (multiple input multiple output) and RF MIMO systems are almost equivalent. It increases the channel capacity of the system almost linearly with the number of transmitting antenna. Optical MIMO transmission with advanced modulation schemes are studied in [136]–[143]. Fig. 9 shows that MIMO systems are more robust to channel fading than SISO or point aperture. It has been observed that when channel state information (CSI) is available at the receiver, an improvement in SNR is directly proportional to the number of transmit and receiver antennae. In case of weak atmospheric turbulence, the outage probability of Gaussian FSO channel is proportional to $[\log(\mathrm{SNR})]^2$ term whereas for moderate to strong turbulence, it is proportional to $[\log(\mathrm{SNR})]$ [138]. Optical MIMO FSO system using $M$-ary Pulse Position Modulation ($M$-PPM) along with multilevel coding (MLC) with low density parity check (LDPC) codes provide excellent coding gain for turbulent regimes. It has shown a drastic coding gain improvement of 57.8 dB at BER = $10^{-6}$ in strong turbulence conditions [144].

FSO MIMO system performs well if the beams are independent or uncorrelated. Otherwise, the performance of FSO system is going to degrade. Another type of diversity that is emerging these days is cooperative diversity [145]–[147] which is a form of distributed spatial diversity that enables multiple terminals to share their resources by cooperative communication so that a virtual antenna array can be built in a distributed fashion. Here, instead of using multiple apertures at the transmitter or receiver end, a single antenna is capable of achieving huge diversity gain. It makes use of neighboring nodes to form virtual antenna array and hence takes the advantage of spatial diversity in distributed manner.

Time diversity with or without codes has also proven to mitigate channel fading in FSO system. This type of diversity is applicable to time selective fading channels which allows repetitive symbols to be transmitted over different coherence time period. If data frame length exceed the channel coherence time, then diversity can be employed by either coding or interleaving. It was observed that in the presence of time diversity, convolutional codes are good choice for weak atmospheric turbulence and Turbo-codes provides a significant coding gain for strong turbulence conditions [148].

- **Adaptive Optics:** Adaptive optics (AO) is used to mitigate the effect of atmospheric turbulence and helps to deliver an undistorted beam through the atmosphere. AO system is basically a closed loop control where the beam is pre-corrected by putting the conjugate of the atmospheric turbulence before transmitting it into the atmosphere [149]–[151]. Increase in transmit power or using diversity can improve the performance of FSO system. But in order have further improvement in SNR with reduce transmit power requirement, AO have proved to be very beneficial. The implementation of AO system in Compensated Earth-Moon-Earth Retro-Reflector Laser Link (CEMRLL) showed significant improvement in the received SNR [152]. AO system makes use of wavefront sensor, wavefront corrector and deformable mirrors either at the transmitter or at the receiver optics to compensate for the phase front fluctuations. Here, a part of the received signal is sent to wavefront sensor that produces a control signal for the actuators of wavefront corrector as shown in Fig. 10. However, a real time wavefront control using conventional AO approach becomes quite difficult for very strong turbulent conditions [151]–[154]. In such situations, non-conventional AO approach is used which is based on the optimization of received SNR or any other system performance metric [155], [156]. Earlier, this non-conventional approach was largely disregarded as it imposed serious limitations for the control bandwidth. But later with the development of high bandwidth wavefront phase controllers e.g., deformable mirrors based on micro-electromechanical systems (MEMS) and with the development of new efficient algorithms, this approach is gaining popularity these days. AO using MEMS for long range FSO system is found in [157], [158].

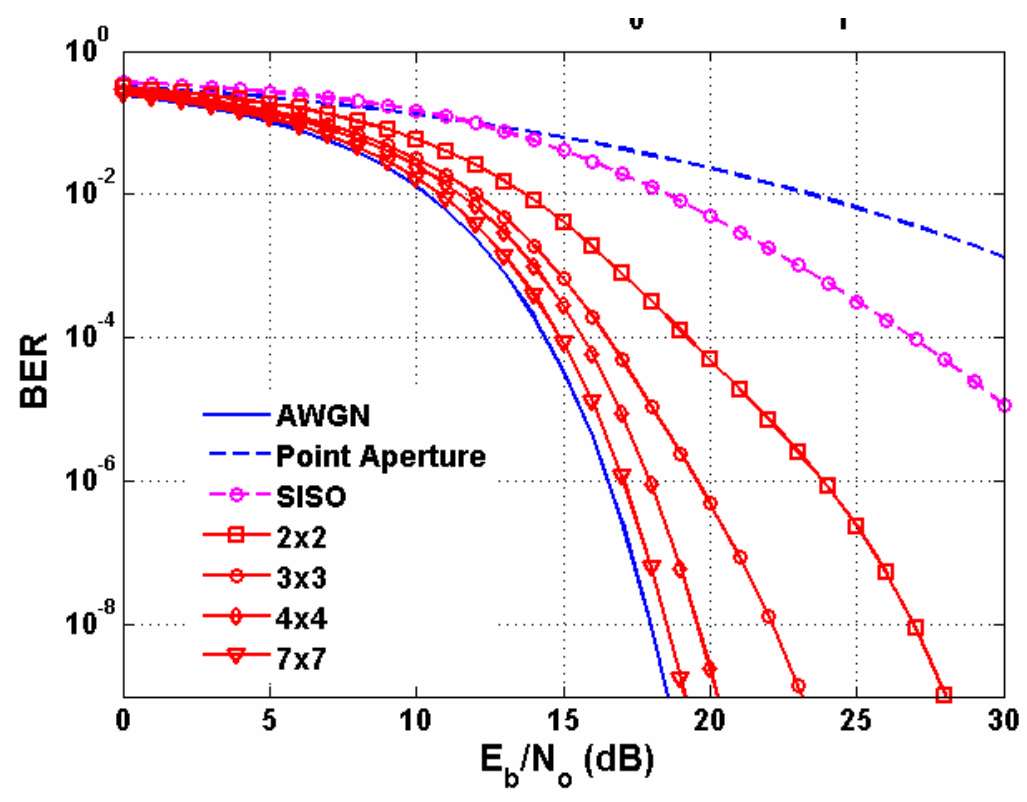

Figure 9. BER performance comparison for SISO and MIMO links at $\sigma_I^2 = 0.8$ and using single aperture diameter = 20 cm [136]

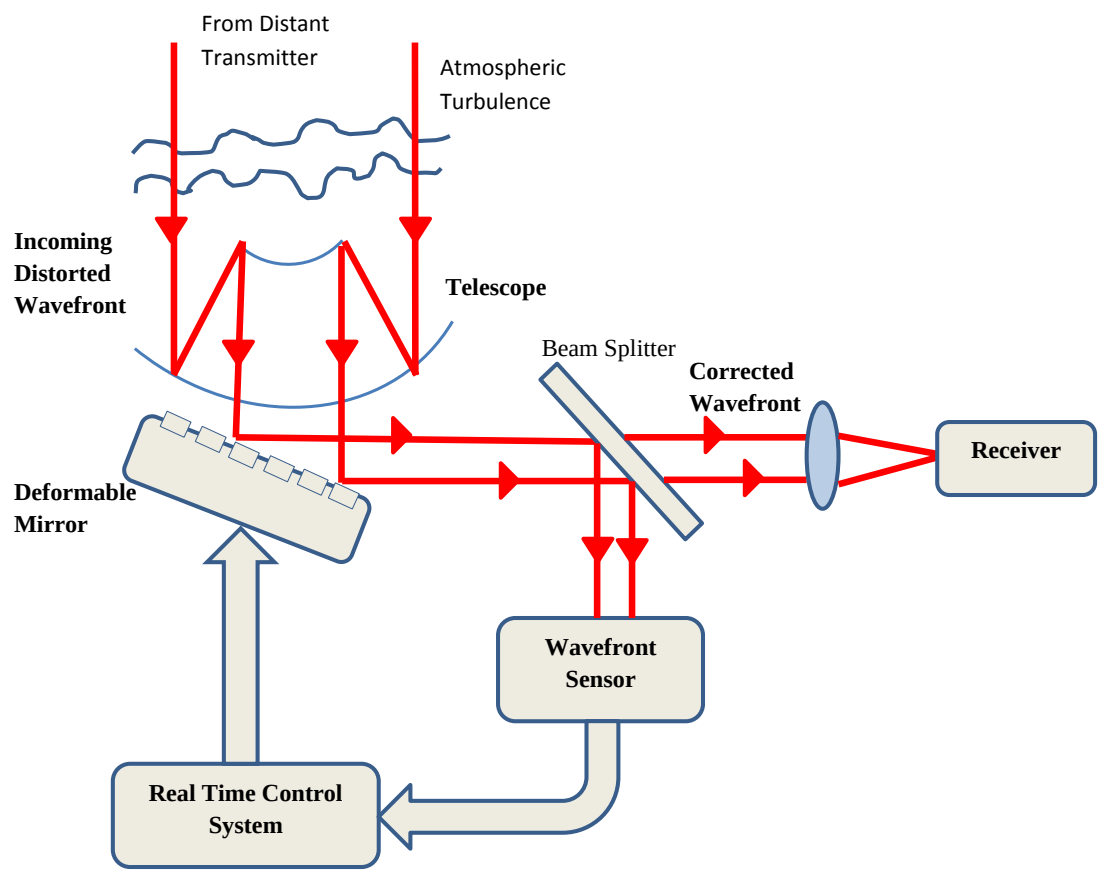

Figure 10. Conventional adaptive optics system

Designing of AO system requires that its closed loop frequency should be at least four times Greenwood frequency [159] (in Hz) given by

$$f_G = \left[ 0.102k^2 \sec\left(\theta\right) \int_{h_o}^{h_o+L} C_n^2 \cdot v_T\left(h\right)^{5/3} dh \right]^{3/5}, \tag{22}$$

where $v_T\left(h\right)$ is the traversal component of wind speed. This frequency tells the speed of AO system to respond to fluctuations due to atmospheric turbulence.

- **Modulation and Coding:** In FSO communication, the choice of modulation and coding schemes depends on two main criteria: optical power efficiency and bandwidth efficiency. Optical power efficiency can be measured by computing optical power gain over On Off Keying (OOK) provided both the modulation schemes have same euclidean distance, $d_{min}$. Power efficient modulation schemes are simpler to implement and are quite effective in mitigating the effect of the turbulence for low data rates. They have to abide with the eye safety regulations, therefore, it limits the maximum propagation distance during high turbulent conditions. Bandwidth efficiency on the other hand, determines maximum data for a given link length with a particular modulation scheme.

In general, FSO communication support variety of binary and multilevel modulation formats. Out of these two format, binary level format is most commonly used due to its simplicity and high power efficiency. Most well known binary modulation schemes are OOK and PPM. OOK modulation scheme requires adaptive threshold in turbulent atmospheric conditions for best results [160], [161]. Due to its simplicity, OOK modulation scheme is very popular in FSO communication system and most commonly it is deployed with IM/DD transmission and receive mechanism. Another detection technique used in OOK modulation scheme is maximum likelihood (ML) detection with perfect CSI [162]. However, due to its implementation complexity, this detection technique didn't gain much popularity. Maximum-likelihood sequence detection (MLSD) can be employed when the receiver is having the knowledge of joint temporal distribution of intensity fluctuations. Other detection techniques [163]–[165] used at the receiver are symbol by symbol maximum likelihood detection, blind detection, VBLAST detection, etc.

In case of $M$-PPM, each symbol interval is divided into $M$ time slots and a non-zero optical pulse is placed at these time slots while other slots are kept vacant. For long distance communication, $M$-PPM scheme is widely used because of its high peak-to-average power ratio (PAPR) that improves its power efficiency [166]. Also, unlike OOK, $M$-PPM does not require adaptive threshold. However, power efficient modulation scheme may not be bandwidth efficient and therefore if the system is bandwidth limited, then multi-level modulation schemes are used. Here, the transmitted data can take multiple amplitude levels and most commonly used multi-level intensity modulation schemes are pulse amplitude modulation (PAM) and quadrature amplitude modulation (QAM) [167], [168]. However, the price paid for bandwidth efficiency is the reduction in power level. Therefore, these modulation schemes are not good choice for high turbulent atmospheric conditions. It is reported in many literatures that in case of high background noise, $M$-PPM is considered to be optimum modulation scheme on Poisson counting channel [169], [170]. With the increase in order of $M$ in $M$-PPM, the robustness against background radiations increases even further due to its low duty cycle and lesser integration interval of photodiode. Owing to various advantages of PPM in FSO communication, various variants of PPM have been developed aiming to enhance the spectral efficiency of the system. These modulation schemes are Differential PPM (DPPM) [171], [172], Differential Amplitude PPM (DAPPM) [173], Pulse Interval Modulation (PIM) [174]. A comparison of bandwidth requirement, PAPR and capacity for all variants of PPM modulation schemes is shown in Table VI.

| Coding | Modulation | Channel Model | Detection Techniques | Reference |
|---|---|---|---|---|
| Convolutional | OOK | Gamma-Gamma | Direct Detection | [198] |
| - | OOK | Log-normal | ML Detection with perfect CSI | [199] |
| - | OOK | Log-normal and Gamma-Gamma | ML sequence detection (MLSD) | [200] |
| Turbo | OOK and PPM | - | Direct Detection | [201] |
| - | OOK | K | Direct Detection | [202] |
| - | OOK | K and IK | Direct Detection | [203] |
| - | OOK | Modified Rician | Coherent Detection | [204] |
| Reed Solomon | PPM and DPSK | Log-normal and Negative exponential | Direct Detection | [205] |
| - | PPM | Gamma-Gamma | Direct Detection | [206] |
| Space Time Trellis code | OOK | Negative exponential and K | Direct Detection | [207] |
| Bit-interleaved coded modulation (BICM) and Multilevel coding (MLC) | PPM | Poisson | Direct Detection | [208] |
| LDPC coded OFDM | OOK, QAM, BPSK, QPSK | Gamma-Gamma | Direct Detection | [194] |
| - | PPM and dual pulse PPM | Negative exponential | Direct Detection | [209] |
| - | DPPM and OOK | Log-normal | Direct Detection | [210] |
| Reed Solomon | PPM | Poisson | Direct Detection | [211] |
| - | SIM DPSK | Gamma-Gamma | Direct Detection | [212] |
| - | SIM BPSK | K | Direct Detection | [213] |
| Block, Convolutional, Turbo codes | OOK | Log-normal | Direct Detection | [214] |
| Turbo coded multi-carrier code division multiple access (MC-CDMA) | SIM-PSK | Gamma-Gamma | Direct Detection | [215] |
| Convolutional code | PPM | Gamma-Gamma | Iterative Detection | [216] |
| LDPC | OOK | Gamma-Gamma | VBLAST-ZF Detection | [217] |
| - | PPM | Gamma-Gamma | VBLAST-ZF Detection | [218] |
| - | Differential pulse position width modulation (DPPM+PWM) | Gaussian | Direct Detection | [219] |
| Hybrid channel (Non-uniform and rate-compatible LDPC codes) & Adaptive Codes | BPSK | Kim model and Gamma-Gamma (Hybrid RF/FSO) | ML Detection | [220] |
| Interleaved concatenated coding (convolutional inner code and a Reed-Solomon outer code) | Binary PPM | Gaussian | Iterative Detection | [221] |

Table VII
LITERATURE ON FSO CODING AND MODULATION SCHEMES USING VARIOUS CHANNEL MODELS AND DETECTION TECHNIQUES

| Modulation Schemes | $M$-PPM | DPPM | DAPPM | DPIM |
|---|---|---|---|---|
| Bandwidth (Hz) | $\frac{MR_b}{\log_2 M}$ | $\frac{(M+1)R_b}{2\log_2 M}$ | $\frac{(M+A)R_b}{2MA}$ | $\frac{(M+3)R_b}{2\log_2 M}$ |
| PAPR | $M$ | $\frac{M+1}{2}$ | $\frac{A(M+1)}{A+1}$ | $\frac{M+1}{2}$ |
| Capacity | $\log_2 M$ | $\frac{2\log_2 M}{M+1}$ | $\frac{2M\log_2(M\cdot A)}{M+A}$ | $\frac{2\log_2 M}{M+3}$ |

Table VI
COMPARISON OF VARIANTS OF PPM MODULATION SCHEME

Optical sub-carrier intensity modulation (SIM) is another modulation format where the base band signal modulates the electrical RF sub-carrier (can be either analog or digital) which is subsequently intensity modulated by the optical carrier. Since sub-carrier signal is sinusoidal signal, therefore a DC bias is added to omit negative amplitude of the transmitted optical signal. SIM does not require adaptive threshold unlike OOK scheme and it is more bandwidth efficient than PPM scheme. Optical SIM inherits the benefits from more mature RF system, therefore, it makes the implementation process simpler. Studies have reported that hybrid PPM-BPSK-SIM gives better results than BPSK-SIM for all levels of atmospheric turbulence [175]. SIM in conjunction with diversity technique improves the BER performance of the FSO system in the presence of atmospheric turbulence. A 4x4 MIMO is considered to be the optimal choice using BPSK-SIM for all turbulent conditions [176]. When this modulation scheme is used with different RF sub-carriers which are frequency multiplexed, then this scheme is know as multiple sub-carrier intensity modulation (MSIM). In this case, each sub-carrier is a narrow band signal and experience less distortion due to inter symbol interference at high data rates. However, the major disadvantage of SIM and MSIM is less power efficiency than OOK or PPM.

Differential Phase Shift Keying (DPSK) has received significant interest due to its power efficiency and 3 dB improvement over OOK modulation [177]–[179]. Since it has reduced power requirement than OOK, DPSK does not have non-linear effects which in turn improves the spectral efficiency of DPSK over OOK modulation. It is reported that sensitivity of DPSK receiver can approach quantum limit theory. However, the cost for implementing DPSK based FSO system is high due to its increased complexity in system design both at transmitter and receiver level.

Error control coding also improves the performance of FSO link by making use of different forward error control (FEC) schemes including Reed-Solomon (RS) codes, Turbo codes, convolutional codes, trellis-coded modulation (TCM) and LDPC. The study of error performance using error correction codes in fading channel has been under research for many years [180]–[182]. These codes add redundant information to the transmitted message so that any kind of error due to channel fading can be detected and corrected at the receiver. The coherence time of FSO system is in the order of milliseconds (about 0.1-10 ms), therefore, the receiver design becomes too complex due to large memories requirement for storing long data frames [183]. On one side, it improves the coding gain of the FSO system but on the other side it leads to delay latencies and complexity of the system. Other attractive option could be interleaving of the transmitted symbols [184]. Since the duration of the fades are random, no single maximum interleaving depth can be used to render the channel completely memory less. Furthermore, interleaving depths that correspond to time separations of 1 ms between successive bits of a code word require the encoder and decoder to contain very large amounts of memory [185]. Coding gain provided by RS and convolutional codes are sufficient in case of weak atmospheric turbulence. The maximum coding gain of convolutional code with constraint length = 3 and code rate = 1/2 for direct detection FSO system using PPM with perfect interleaving is 7 dB for clear weather conditions and 11 dB in moderate turbulent conditions. The use of soft decision Viterbi decoding in this case provided significant improvement in BER of the system even when the interleaving depths were insufficient to render the channel memory less [186]. RS codes provide good coding gain when implemented with PPM. The performance of RS (255, 127) coded PPM provides coding gain of 6 dB [187]. The improvement in RS codes increases with the increase in block length. RS coded PPM (63, 37, 64) at code rate = 3/5 matches the performance of 64-PPM, but RS (262143, 157285, 64) gives better performance at BER = $10^{-6}$. In case of strong atmospheric turbulence, Turbo, Trellis or LDPC codes are preferred. Turbo codes can be arranged in any of the three different configurations- parallel concatenated convolutional codes, serial concatenated convolutional codes and hybrid concatenated convolutional codes. Parallel concatenated convolutional codes are most popularly used in which two or more constituent systematic recursive convolutional encoders are linked through an interleaver. For very high data rate transmission, LDPC codes are preferred over Turbo codes due to its reduced decoding complexity and computational time. Variable rate LDPC codes can further increase the channel capacity and provide good coding gain [188], [189]. Various analysis has been carried out for the use of LDPC codes in MIMO FSO system [190], [191]. It was observed that LDPC coded MIMO FSO system using $M$-PPM provides better performance over uncoded system in case of strong atmospheric turbulence and large background noise set to -170 dBJ. A coding gain of 10-20 dB was observed over uncoded system at BER = $10^{-12}$ [192], [193]. Also, bit interleaved coded modulation (BICM) scheme proposed by I.B Djordjevic [192] provide excellent coding gain when used with LDPC coded FSO system. Orthogonal frequency division multiplexing (OFDM) combined with suitable error control coding is also considered a very good modulation format for improving BER performance of FSO IM/DD systems [194].

At the receiver, various efficient decoding algorithms have been proposed to decode the generated codes. Theoretically, ML decoding at the receiver can provide better data recovery but its usage is limited due to implementation complexity. Symbol-by-symbol maximum a posterior (MAP) decoding algorithm is computationally complex and is not a preferred choice for implemention on VLSI chip. However, logarithmic version of the MAP (log-MAP) algorithm [195] and the Soft Output Viterbi Algorithm (SOVA) [196] are the practical decoding algorithms for implementation using Turbo codes. Out of these two, log-MAP algorithm gives the best performance but it is computationally very complex. Simplified-log-MAP algorithm performs very close to the log-MAP and is less complex aswell [197]. Some of decoding algorithms used in LDPC codes are belief propagation and message passing. VBLAST-ZF (Vertical Bell Labs Layered Space Time Zero Forcing) detection algorithm is suggested in LDPC coded MIMO FSO system for reducing the decoding complexities at the receiver. A summary of literature pertaining to various coding and modulation schemes using various channel models and detection techniques is presented in Table VII.

- **Background Noise Rejection:** The major source of background noise is due to day time solar radiations. This can be mitigated with the help of spatial filters along with suitable modulation technique that has high peak-to-average power [53], [222]. The most suitable modulation scheme is $M$-PPM to combat the effect of solar background noise radiations (as the noise is directly proportional to slot width). High order PPM scheme is reported as potential modulation scheme for inter-satellite links as it is more power efficient and drastically reduce the solar background noise [223]. However, $M$-PPM is not suitable for bandwidth limited system. In that case, DPIM is a preferred choice as it does not require synchronization and is both capacity and bandwidth efficient [89], [224]. Designing receiver with narrow FOV and choosing filters with spectral width less than 1 nm is another approach to reduce the background noise [225]. It has been shown that adaptive optics and deformable mirrors comprising of array of actuators can result in significant improvement due to background noise by reducing the receiver FOV. Analysis presented in [226] shows that inter-planetary FSO link between Earth and Mars has achieved 8.5 dB improvement in extreme background and turbulent conditions using adaptive optics and array of actuators (hundred 1 meter telescope) with PPM modulation scheme. During moderate background conditions, the improvement was decreased to 5.6 dB. Another analysis presented in [227] gives the performance improvement of 6 dB by using array of 900 actuators and adaptive optic technique with 16 PPM modulation scheme.
- **Hybrid RF/FSO:** The performance of FSO communication is drastically affected by weather conditions and atmospheric turbulence. This can lead to link failures or poor BER performance of FSO system. Therefore, in order to improve the reliability and improve the availability of the link, it is wise to pair up FSO system with a more reliable RF system. Such systems are called hybrid RF/FSO and are capable of providing high link availability even in adverse weather conditions [69]. The major cause of signal degradation in RF transmission is due to rain (as the carrier wavelength is comparable to the size of the rain drop) and in FSO communication is due to fog. So, the overall system availability can be improved by using low data rate RF link as a back up when FSO link is down. In [228], the availability of an airborne hybrid RF/FSO link is evaluated. It was observed that the FSO link provides poor availability during low clouds conditions due to the attenuation by cloud particles and temporal dispersion. However, a significant improvement was observed when a hybrid RF/FSO link was used as RF signals are immune to cloud interference. The conventional approach of hybrid RF/FSO causes inefficient use of channel bandwidth [229]. Also, a continuous switching between RF and FSO system could bring down the entire system. Therefore, a new approach as suggested in [230] gives symbol rate adaptive joint coding scheme wherein both RF and FSO subsystem are active simultaneously and saves channel bandwidth. Hybrid channel coding is also capable of utilizing both the links by combining non-uniform codes and rate adaptive codes where their code rates are varied according to the channel conditions [220].

  Hybrid RF/FSO link provides great application in mobile ad hoc networks (MANETs) [231]. A reconfigurable networking environment can be formed in MANETs by combination of wireless sensor network (WSN) technology and mobile robotics. However, the performance of this network is limited by the per node throughput provided by RF based communication. Therefore, the combination of RF and FSO provides tremendous increase in per node throughput of MANETs. The implementation of hybrid RF/FSO MANET with real-time video data routing across 100 Mbps optical link and 802 .11g RF transceiver has been studied in [232].

  The RF wireless network poses a strong limitation on its capacity and throughput owing to growing development in communication technology [233]. With the increasing number of users, the chances of interference from the neighboring nodes increases and that limits the performance of the RF system. FSO system on the other hand is highly directional and has very narrow beam divergence. This makes FSO system immune from any kind of interference. Therefore, the combination of FSO and RF can help in solving the capacity scarcity problem in RF networks. The throughput and capacity of hybrid RF/FSO link is given in [234]–[236]

### B. TCP Upper Layer Methods

There has been a lot of research on the performance mitigation of atmospheric turbulence in physical layer. For the last few years, researchers have gained attention to work on modeling and performance evaluation of upper layers including link layer, transport layer, application layer in order to improve the performance of FSO communication [237], [238]. In addition to the physical layer methods, various techniques like re-transmission, re-routing, cross connection between different layers, delay tolerant networking, etc., are used to improve the performance of FSO in all weather conditions [239]–[241].

- **Re-transmission:** A re-transmission protocol such as automatic repeat request (ARQ) is widely used in data communication for reliable data transfer. Here, the transmission is carried out in the form of packets of certain frame length. If due to some reason, the receiver does not acknowledge the transmitted packet within speculated time frame, then the packet is re-transmitted. This process repeats till a positive acknowledgment is received by the transmitter from the receiver or the preset counter value is exceeded. So, this kind of stop, wait and go-back-$N$ ARQ scheme results in huge delay, large energy consumption and bandwidth penalties due to re-transmission process [242], [243]. Therefore, another variant of ARQ is selective repeat ARQ (SR-ARQ) in which data packets are continuously transmitted from transmitter to the receiver without the need to wait for individual acknowledgment from the receiver. The receiver will continue to accept and acknowledge the received frame. If any frame is not acknowledged after certain period, it is assumed to be lost and re-transmitted. ARQ protocol can be implemented either at data link layer or at transport layer [244]. In either case, the receiver terminal must have sufficient data storage capability to buffer the received data at least for the time period specified by window size. In [241], [245], the performance of FSO system in weak atmospheric turbulence has been studied when the ARQ and FEC schemes are used in link layer. Lee and Chan in [246] examined the performance of transmission control protocol (TCP) and observed very poor throughput even with 10 dB of link margin and 16 diversity transmitters/receivers over clear weather conditions. The performance of TCP over FSO channel as shown in Fig. 11 has been analyzed when SR-ARQ scheme is used for the link layer in [247]. Here, TCP-Reno and three-dimensional (3-D) Markov chain model that includes the exponential back-off phase were used for modeling the TCP operation. The throughput of TCP was analyzed and it was observed that the atmospheric turbulence has severe impact on TCP [246] and use of SR-ARQ in link layer can significantly maximize the throughput of TCP.

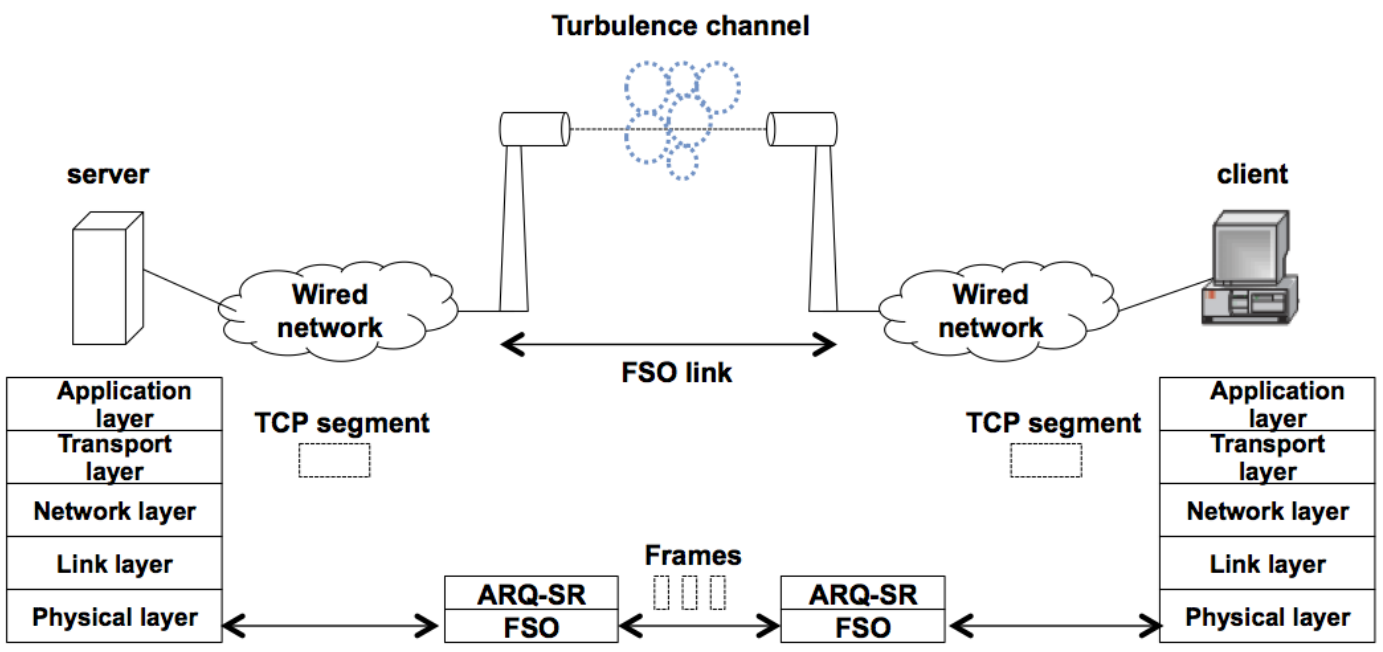

Figure 11. TCP connection over FSO link when used with SR-ARQ [247]

Another variant of ARQ that has been studied by various researcher is hybrid ARQ (HARQ) which uses a combination of FEC coding and ARQ error control [248]–[250]. The outage probability of different HARQ scheme in the strong turbulence regimes has been investigated and it was found that good performance gain can be achieved using this scheme [251]. However, this scheme has large bandwidth penalties and delay latencies. Recently, combination of cooperative diversity with ARQ (CARQ) has gained interest that has shown remarkable results for combating turbulence induced fading in FSO channel. A modified version of CARQ with lesser transmission delays and improved energy consumption is M-CARQ. This modified scheme allows relay nodes to store a copy of frames for more efficient response to transmission failure due to atmospheric turbulence [252], [253].

Another protocol i.e., Rateless Round Robin protocol is used in FSO networks which is based on the combination of channel coding (FEC) and packet level coding (a form of ARQ) [254], [255]. FEC is applied to the transmitted data and reverse link acknowledgment. Cyclic redundancy check (CRC) is used to verify the integrity of the received packets after FEC decoding [256]. It is observed that Rateless Round Robin is an effective error control design for practical FSO applications even during very strong turbulence when the channel availability is less than 45% [257].

- **Reconfiguration and re-routing:** Path reconfiguration and data re-routing is carried out in order to increase the availability and reliability of the FSO link during loss of LOS or adverse atmospheric conditions or device failure. Through dynamic reconfiguration of the nodes in FSO network using physical and logical control mechanism, link availability is improved drastically. In physical layer, reconfiguration is achieved using pointing, acquisition and tracking (PAT) and in logical layer, it makes use of autonomous reconfiguration algorithms and heuristics. Here, the data packets are re-routed through other existing links that could be either an optical link or low data rate RF link. Various topology control mechanism and re-routing algorithms have been investigated for FSO network [258]–[262]. Autonomous topology control and beam reconfiguration is achieved through: (a) the topology discovery and monitoring process, (b) the decision making process by which a topology change has to be made, (c) the dynamic and autonomous re-direction of beams (based on algorithms) to new receiver nodes in the network, and (d) the dynamic control of these beams for link re-direction [259]. Therefore, reconfiguration and re-routing improves the reliability of FSO link but at the cost of huge processing delays. A good design engineer has to ensure the restoration of link through reconfigurability without significant impact in delay and at reduce cost. For this, routing protocol should be designed in such a way so that during re-routing process, the path which has minimum delay or least number of hops should be given priority. Sometimes, all the routing routes are computed prior to their actual need and are stored in routing tables. Such type of routing is classified as ‘proactive routing’ protocol. However, this routing protocol is not suitable for large networks as it imposes high overhead to the network and that makes it bandwidth inefficient. Examples of proactive routing protocol are: Destination sequenced distance-vector (DSDV) routing protocol [263], optimized link state (OLS) routing protocol [264], wireless routing protocol (WRP) [265]. Another routing protocol that generates very less overhead as compare to proactive routing is called 'reactive routing' and computes new routes only 'on demand'. A new route is established only during the failure of the existing route. However, this leads to prolonged latency in data delivery. Examples of reactive routing protocols are: Ad hoc on-demand (AODV) routing [266], location aided routing (LAR) [267], temporary ordered routing (TOR) protocol and dynamic source routing (DSR) [268]. Combination of both proactive and reactive routing protocols is called 'hybrid routing' protocol. It divides the network into clusters and apply proactive route updates within each cluster and reactive routing across different clusters. Examples of hybrid routing protocol are zone routing protocol (ZRP) [269] and Ad hoc on-demand distance vector hybrid (AODVH) [270]. A comparison of some of popular routing protocols that can be implemented in wireless network are presented in Table VIII. Further, the routing protocols can be classified as negotiation-based [271], multipath-based [272], [273], query-based [274], [275], QoS-based [276], [277], or coherent-based routing [278] as shown in Fig. 12.

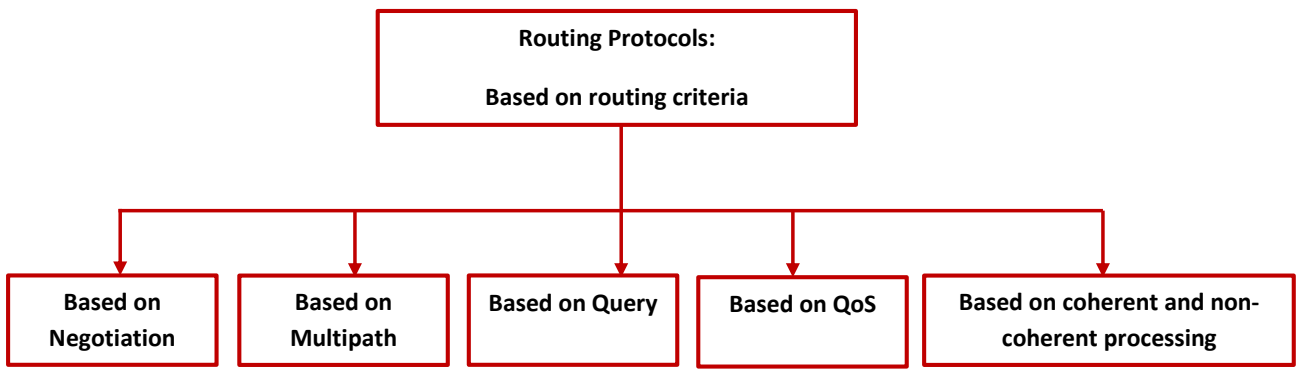

Figure 12. Classification of routing protocols

- **Quality of Service Control:** The QoS in FSO communication is measured in terms of data rate, latency, delay jitter, data loss, energy consumption, reliability and throughput efficiency. The data transfer from one node to another in FSO communication system should meet the given requirements of special QoS class otherwise the offered services can not be used by end-users in a satisfying way. For this, the main challenge in FSO network is to optimize the performance of communication system measured in (a) end-to-end connection delay, (b) delay variation, (c) packet rejection rate, and (d) overhead [90]. It is suggested in some literatures that modification in different layers improve the QoS of FSO system. Various methods that address application level QoS control are: application admission control algorithm [292], multipath and multi-speed (MMSPEED) routing protocols proposed in [293], tunneling [294], DSR protocol [269], etc. A routing algorithm that improves the QoS of network and medium access control (MAC) layer is proposed in [295]. The MAC layer QoS can be classified into (a) channel access policies, (b) scheduling and buffer management, and (c) error control. This routing protocol provides energy efficient real time FSO communication. Some of the QoS routing protocols implemented in wireless networks are: multipath routing protocol (MRP) [296], sequential assignment routing (SAR) protocol

| Protocol | Classification | Energy Consumption | Route Overhead | Scalability |
|---|---|---|---|---|
| Low Energy Adaptive Clustering Hierarchy (LEACH) [279] | Hierarchical | High | High | Good |
| Geographic Adaptive Fidelity (GAF) [280] | Location | Low | Average | Good |
| Geographical and Energy Aware Routing (GEAR) [281] | Location | Low | Average | Limited |
| Directed Diffusion (DD) [282] | Flat | Low | Low | Limited |
| Rumor Routing (RR) [283] | Flat | Low | Low | Good |
| Gradient Based Routing (GBR) [284] | Flat | Low | Low | Limited |
| Adaptive Threshold Sensitive Energy Efficient Network (ATEEN) [285] | Hierarchical | High | High | Good |
| Dynamic Source Routing (DSR) [286] | Reactive | Average | Average | Limited |
| Location Aided Routing (LAR) [287] | Reactive | Low | Low | Limited |
| Link Quality Source Routing (LQSR) [288] | Reactive | Low | Low | Limited |
| Temporally Ordered Routing Algorithm (TORA) [289] | Reactive | Low | Average | High |
| Zone Routing Protocol (ZRP) [290] | Hybrid | Low | Low | High |
| Ad Hoc On Demand Distance Vector (AODV) [291] | Reactive | High | High | Limited |

Table VIII
COMPARISON OF REAL TIME ROUTING PROTOCOLS

[297], energy-aware QoS routing protocol [298], SPEED [299], multi constrained QoS multi-path routing (MCMP) protocol [300], QoS-based energy-efficient sensor routing (QuESt) protocol [301], etc.

- **Others:** Re-playing is another technique to promote end-to-end connectivity of FSO link. If re-routing or re-transmission is not possible, then FSO network will replay up to 5 sec of data from the edge node [302]. Delay (or disruption) tolerant networking (DTN) technique is applied for the networks with intermittent connectivity and therefore, it is a good candidate for FSO communication having extreme atmospheric conditions [303], [304].

## IV. Orbital Angular Momentum for FSO System

Angular momentum is one of the most fundamental physical quantity in both classical and quantum mechanics. It is classified as spin angular momentum (SAM) and orbital angular momentum (OAM). SAM is associated with the spin of the photon and thus, it is related with polarization. On the other hand, OAM is associated with the helicity photon wavefront and therefore, it is related to spatial distribution. It was reported in [305] that a beam having helical shaped phase front described by azimuthal phase term exp($il\theta$) carries OAM of $l\bar{h}$ per photon where $l$ is topological charge with any integer value, $\theta$ the azimuth angle and $\bar{h}$ the Plank's constant $h$ divided by $2\pi$. Therefore, unlike SAM, which can take only two possible states of $\pm\bar{h}$, the OAM can have infinite number of states corresponding to different values of $l$. In principle, infinite number of bits is carried by OAM of single photon. This makes OAM a potential candidate for high capacity communication systems. Also, orthogonality among beams with different OAM states allow additional degree of freedom by multiplexing of information carrying OAM beams. The possibility to generate and analyze states with different OAM by using interferometric or holographic methods [306]–[308] permit the development of energy-efficient FSO communication systems. Further, OAM based FSO system has provided good performance in atmospheric turbulence when used with suitable encoding or modulation format or adaptive optics system.

An OAM beam is formed by attaching azimuthal phase term to the Gaussian beam as $U\left(r,\theta\right) = A\left(r\right) \cdot \exp\left(il\theta\right)$. Here, $A\left(r\right)$ is the amplitude at the waist of the Gaussian beam and $r$ the radial distance from the center axis of the Gaussian beam. When data is encoded to OAM beam, it is expressed as $U\left(r,\theta,t\right) = S\left(t\right) \cdot A\left(r\right) \cdot \exp\left(il\theta\right)$ where $S\left(t\right)$ is the data to be transmitted. With multiplexing of $N$ information carrying OAM beams, the resultant field is expressed as $U_{\mathrm{Mux}}\left(r,\theta,t\right) = \sum_{m=1}^{N} S_m\left(t\right) \cdot A_m\left(r\right) \cdot \exp\left(il_m\theta\right)$. It is to be noted that OAM of individual beam is not modified when propagated through free space or spherical lenses. For de-multiplexing OAM beams, an inverse of azimuthal phase term $\exp\left[i\left(-l_n\right)\theta\right]$ is used and received de-multiplexed beam is then given as

$$U_{\mathrm{Rx}}\left(r,\theta,t\right) = \exp\left[i\left(-l_n\right)\theta\right] \cdot \sum_{m=1}^{N} S_m\left(t\right) \cdot A_m\left(r\right) \cdot \exp\left(il_m\theta\right). \tag{23}$$

The capacity and spectral efficiency of FSO links using OAM is increased by employing several techniques like: (i) combining multiple beams with different OAM values, (ii) using positive or negative OAM values, (iii) using wavelength division multiplexing (WDM), (iv) polarization multiplexing, or (v) using two groups of concentric rings. Recent reports in [309] have demonstrated 2.56 Tbps data rate transmission with spectral efficiency of 95.7 bps/Hz using four light beams with 32 OAM modes employing 16 quadrature amplitude modulation (QAM). In [310], 100.8 Tbps data transmission has been reported using 42 wavelengths with 24 modes. This shows that OAM has tremendous potential for increasing the capacity of FSO system. However, these high capacity transmissions were limited to short transmission distances only where the effect of turbulence was not considered. Therefore, OAM based FSO system is capable of delivering huge data returns in inter-satellite links or deep space mission where atmospheric turbulence does not pose any problem.

It has been reported that OAM beams are highly sensitive even in case of weak atmospheric turbulence due to redistribution of energy among various OAM states leading to time varying crosstalk [311]. An OAM beam has a doughnut shape with less power and large phase fluctuations in the center. Since the orthogonality of the beam is dependent upon helical phase front, therefore practical implementation of OAM based FSO system in the presence of atmospheric turbulence seems to be challenging. Single information carrying OAM state result in random and bursty error in the presence of atmospheric turbulence. In case of multiplexed data channels having different OAM values, the crosstalk among adjacent channels degrade the performance of the system [312]. In [313], experimental investigation was carried out for OAM based multiplexed FSO communication link through emulated atmospheric turbulence. The results indicated that turbulence induced signal fading and crosstalk could significantly deteriorate the link performance. However, very recently researchers have reported that the effect of atmospheric turbulence is mitigated in OAM based FSO system by using suitable channel coding and wavefront correction techniques. Channel codes are used to correct the random errors caused by atmospheric turbulence for single OAM state and wavefront correctors take care of cross talk among adjacent OAM states. Various channel coding techniques like RS codes in [314], LDPC codes in [315] have proved quite beneficial in improving the performance of OAM based FSO system. Other techniques like holographic ghost imaging system [316], adaptive optics [317], [318] help in controlling OAM crosstalk and thus is capable of providing high data rates even in adverse atmospheric conditions. Combination of error correcting codes and wavefront correction methods have also shown good results using OAM based FSO system in the presence of atmospheric turbulence. In [319], RS codes in combination with Shark-Hartmann wavefront correction method has been adopted to combat the detrimental effect of atmospheric turbulence. It was shown that the effect of atmospheric turbulence was reduced and satisfactory BER performance was achieved even during strong atmospheric turbulence. Fig. 13 shows the block diagram representation of OAM based FSO multiplexed system using channel coding technique. The transmitter consists encoder, modulator, laser source, computer controlled hologram along with spatial light modulator (SLM) and multiplexer. The data from different independent sources are encoded, modulated and transformed into OAM beams by adding a spiral phase mask with different charges ($l$). These OAM beams are then multiplexed together and sent onto FSO channel. The multiplexing of OAM beams is considered as a form of spatial multiplexing which is capable of enhancing the capacity and spectral efficiency of the FSO link. At the receiver side, OAM beams are de-multiplexed and an inverse spiral phase mask with charge ($-l$) is used to remove the azimuthal phase term exp ($il\theta$) of OAM beam to recover the plane phase front of the beam. This optical beam is then passed through detector, demodulator and decoder to recover information data.

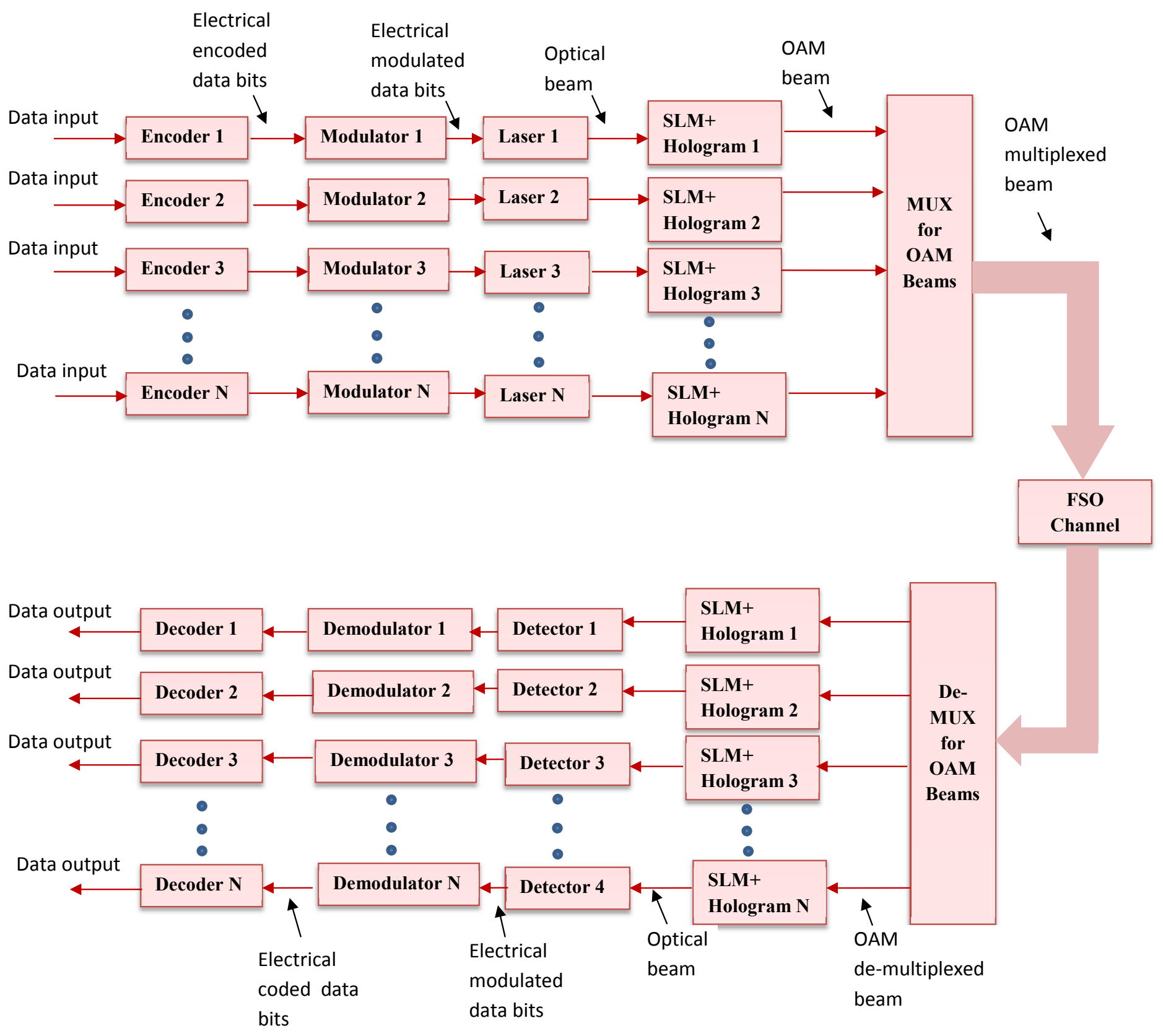

Figure 13. FSO communication system using multiplexed OAM beams and channel coding

## V. Future Scope of FSO Communication

FSO communication seems to have promising future as it provides a cost effective connectivity alternatives for several applications like: last mile access, cell cite back haul for mobile networks, fiber backup and much more. The FSO communication has experienced a rapid growth in the last few years despite of various crisis in the global market. This technology has demonstrated less capital expenditure with huge returns in very little time due to (i) easy availability of components, (ii) quick deployment (as it does not seek permission from municipal corporation for digging up of street), and (iii) no licensing fee required. FSO communication provides a back up protection for fiber based system in case of accidental fiber damage. Also, FSO technology is used to provide high speed data connectivity for distance ranging from few cms ( like in optical interconnect networks) up to few meters/kms (like in wireless local area networks (WLAN), metro area extensions, wireless body area network (WBAN), etc.). This technology has proved to outreach the capacity of RF wireless link by providing 10 Gbps wireless optical link. With the availability of 10G Ethernet switches in the market these days, FSO technology is capable of providing promising gigabit Ethernet access for high rise network enterprise or bandwidth intensive applications (e.g., medical imaging, HDTV, hospitals for transferring large digital imaging files or telecommunication) or intra campus connections. FSO technology provides good solution for cellular carriers using 4G technology to cater their large bandwidth and multimedia requirement by providing a back haul connection between cell towers. Using ultra short pulse (USP) laser, FSO communication provide up to 10 Gbps back haul connection without deploying fiber cables. It is believed that FSO technology is the ultimate solution for providing high capacity last mile connectivity up to residential access. Instead of hybrid fiber-coax systems, hybrid fiber-FSO system may cater the high bandwidth and high data rate requirements of end users.

FSO technology allows connectivity to remote places where physical access to 3G or 4G signals is difficult. It involve integration of terrestrial and space networks with the help of High Altitude Platforms (HAPs)/UAVs by providing last mile connectivity to sensitive areas (e.g., disaster relief, battlefields, etc.) where high bandwidth and accessibility are necessary. An upcoming project by Facebook is an example that will allow internet access to sub-urban or remote areas by providing aerial connectivity to users using FSO link. It is proposed that for sub-urban areas in limited geographical regions, solar-powered high altitude drones will be used to deliver reliable internet connections via FSO links. For places where deployment of drones is uneconomical or impractical (like in low population density areas), LEO and GEO satellites can be used to provide internet access to the ground using FSO link. Hybrid RF/FSO communication has been proposed for wireless sensor network due to their low energy consumption requirement.

FSO interconnects (FSOI) over very short distances like chip-to-chip or board-to-board have gained popularity these days as it potentially addresses complex communication requirement in optoelectronic devices. FSOI technology offers the potential to build interconnection networks with higher speed, lower power dissipation and more compact packages than possible with electronic VLSI technology. However, the cost of optoelectronic devices, their integration and overall packaging makes FSOI a costly affair. A throughput upto 1 Tbps per printed circuit board (PCB) board has been experimentally demonstrated in [320] using 1000 channels per PCB with 1 mm optical beam array at 1 Gbps per channel.

FSO technology when used over a mobile platform can be deployed in armed forces as it demands secure transmission of information on the battlefield. Intelligence, Surveillance, and Reconnaissance (ISR) platforms can deploy this technology as they require to disseminate large amount of images and videos to the fighting forces, mostly in real time. Besides, this technology can be a good alternative to acoustic and tethered underwater communications for short distances. It can be used for various applications like undersea explorations, monitoring ocean currents and winds for improving weather forecast, providing tsunami warning by measuring seismic activities, etc. It can be used by navy surface ships for communicating with underwater submarines. The acoustic technology that is currently being used is not suitable for high data rates in real time environment. However, FSO underwater technology is capable of providing high data rates in real time applications. This technology can also be used with underwater sensor nodes that collects long term geo-physical data. At present, various devices are recovered to off load their data before they are deployed again, which is a resource intensive task. In the future, these underwater sensor nodes can use FSO technology to wirelessly off load their data to an interrogating underwater vehicle equipped with an optical modem.

## VI. Conclusions

The tremendous growth in the number of multimedia users and internet traffic in the recent years has incurred a substantial strain on RF system operating at low data rates. Due to this huge explosion in information technology that is driving the information business to higher and higher data rates, there is a need to switch from RF domain to optical domain. FSO communication is capable of providing LOS wireless connection between remotes sites with very high bandwidths. This technology is considered to be the promising technology in near future which can meet very high speed and huge capacity requirements of current day communication market. However, in order to fully utilize the terabit capacity of FSO system, it has to overcome various challenges offered by heterogeneous nature of atmospheric channel. FSO system is vulnerable towards various atmospheric phenomenon like absorption, scattering, atmospheric turbulence and adverse weather conditions. Various techniques implemented either at physical layer or at network layer help to combat the detrimental effect of atmosphere on the quality of the laser beam. Several fading mitigation techniques that were initially proposed for RF works well for FSO communication aswell e.g., diversity, adaptive optics, error control codes, modulation, etc. Besides this, the complementary nature of RF and FSO has motivated the design of hybrid RF/FSO system which ensure carrier class availability for almost all weather conditions. Also, modifications in the upper layers of TCP model like application, transport and link layer with suitable protocols and algorithms help in improving the reliability of FSO system.

Hence, it is clear that after so much advancement in FSO communication, this technology seems to have very high growth prospects in the near future. Many commercial product for FSO terrestrial and space links are already available in market and hopefully very soon this technology will bring worldwide telecommunication revolution.

## References

[1] V. W. S. Chan, "Free-space optical communications," *J. Lightwave Tech.*, vol. 24, no. 12, pp. 4750–4762, 2006.

[2] F. R. Gfeller and U. H. Bapst, "Wireless in-house data communication via diffuse infrared radiation," *Proc. IEEE*, vol. 67, no. 11, pp. 1474–1486, 1979.

[3] A. J. C. Moreira, A. M. Tavares, R. T. Valadas, and A. M. de Oliveira Duarte, "Modulation methods for wireless infrared transmission systems performance under ambient light noise and interference," *Proc. SPIE, Wireless Data Trans.*, vol. 2601, pp. 226–237, 1995.

[4] A. M. Street, P. N. Stavrinou, D. C. O ' B rien, and D. J. Edward, "Indoor optical wireless systems- A review," *Opt. and Quant. Electr.*, vol. 29, pp. 349–378, 1997.

[5] A. P. Tang, J. M. Kahn, and K. P. Ho, "Wireless infrared communication links using multi-beam transmitters and imaging receivers," in *Proc. IEEE Int. Conf. on Commun., Dallas, U.S.A.*, pp. 180–186, 1996.

[6] J. B. Carruthers and J. M. Kahn, "Angle diversity for nondirected wireless infrared communication," *IEEE Trans. Comm.*, vol. 48, no. 6, pp. 960–969, 2000.

[7] R. Ramirez-Iniguez and R. J. Green, "Indoor optical wireless communications," in *IEE Colloquium on Opt. Wireless Comm.*, vol. 128, pp. 14/1–14/7, 1999.

[8] D. J. T. Heatley, D. R. Wisely, I. Neild, and P. Cochrane, "Optical Wireless: The story so far," *IEEE Comm. Mag.*, vol. 36, no. 12, pp. 72–74, 79–82, Dec. 1998.

[9] J. Fernandes, P. A. Watson, and J. Neves, "Wireless LANs: Physical properties of infrared systems vs mmw systems," *IEEE Comm. Mag.*, vol. 32, no. 8, pp. 68–73, Aug. 1994.

[10] Z. Ghassemlooy and W. O. Popoola, *Terrestial Free-Space Optical Communications*, ch. 17, pp. 356–392. InTech, 2010.

[11] H. T. Yura and W. G. McKinley, "Optical scintillation statistics for IR ground-to-space laser communication systems," *Appl. Opt.*, vol. 22, no. 21, pp. 3353–3358, Nov. 1983.

[12] V. Sharma and N. Kumar, "Improved analysis of 2.5 Gbps-inter-satellite link (ISL) in inter-satellite optical wireless communication (ISOWC) system," *Opt. Comm.*, vol. 286, pp. 99–102, 2014.

[13] D. Hochfelder, "Alexander graham bell," Encylopedia Britannica, 2015.

[14] D. L. Hutt, K. J. Snell, and P. A. Belanger, "Alexander graham bell's photophone," Tech. Report: Optic & Photonics News, 1993.

[15] J. Hecht, "Beam: The race to make the laser," Tech. Report: Optics & Photonics News, 2005.

[16] F. Bellinne and D. E. Tonini, "Flight testing and evaluation of airborne multisensor display systems," *J. Aircraft*, vol. 7, no. 1, pp. 27–31, 1970.

[17] R. Deadrick and W. F. Deckelman, "Laser crosslink subsystem - An overview," in *Proc. SPIE, Free Space Laser Comm. Tech. IV*, vol. 1635, (Los Angeles, CA), pp. 225–235, 1992.

[18] M. Jeganathan, K. E. Wilson, and J. R. Lesh, "Preliminary analysis of fluctuations in the received uplink-beacon-power data obtained from the GOLD experiments," TDA Progress Report 42-124, Comm. Sys. and Research Sec., pp. 20-32, 1996.

[19] K. E. Wilson, "An overview of the GOLD experiment between the ETS-VI satellite and the table mountain facility," TDA Progress Report 42-124, Comm. Sys. and Research Sec., pp. 9-19, 1996.

[20] J. Horwath, M. Knapek, B. Epple, M. Brechtelsbauer, and B. Wilkerson, "Broadband backhaul communication for stratospheric platforms: The stratospheric optical payload experiment (STROPEX)," in *Proc. SPIE, Free-Space Laser Comm. VI*, vol. 6304, (San Diego, California, USA), Sept. 2006.

[21] A. Biswas, D. Boroson, and B. Edwards, "Mars laser communication demonstration: What it would have been," *Proc. SPIE, Free Space Laser Comm. Tech. XVIII*, vol. 6105, 2006.

[22] V. Cazaubiel, G. Planche, V. Chorvalli, L. Hors, B. Roy, E. Giraud, L. Vaillon, F. Carré, and E. Decourbey, "LOLA: A 40,000 km optical link between an aircraft and a geostationary satellite," in *Proc. 6th International Conf. Space Optics*, (Netherlands), Jun. 2006.

[23] "Laser communications relay demonstration: The next step in optical communications," Goddard Space Flight Center, (NASA), Tech. Report: Weblink: *http://www.nasa.gov/pdf/742122main_LCRDFactSheet3.pdf*.

[24] Weblink: *http://artolink.com*.

[25] Weblink: *http://www.fsona.com* .

[26] R. M. Sova, J. E. Sluz, D. W. Young, J. J. C., A. Dwivedi, N. M. Demidovich III, J. E. GrJohnaves, M. Northcott, J. Douglass, J. Phillips, D. Don, A. McClarin, and D. Abelson, "80 Gb/s free-space optical communication demonstration between an aerostat and a ground terminal," in *Proc. SPIE, Free Space Laser Comm. VI*, vol. 6304, 2006.

[27] W. D. Williams, M. Collins, D. M. Boroson, J. Lesh, A. Biswas, R. Orr, L. Schuchman, and O. Scott Sands, "RF and optical communications: A comparison of high data rate returns from deep space in the 2020 timeframe," Tech. Report: NASA/TM-2007-214459, 2007.

[28] J. H. Franz and V. K. Jain, *Optical Communications: Components and Systems: Analysis, Design, Optimization, Application*. Narosa Publishing House, 2000.

[29] H. Henniger and O. Wilfert, "An introduction to free-space optical communications," *J. Radioeng.*, vol. 19, no. 2, pp. 203–212, 2010.

[30] M. Kaine-Krolak and M. E. Novak, "An introduction to Infrared technology: Applications in the home, classroom, workplace, and beyond," (Trace R&D Center, University of Wisconsin, Madison, WI), 1995.

[31] V. G. Sidorovich, "Solar background effects in wireless optical communications," *Proc. SPIE , Opt. Wireless Comm. V*, vol. 4873, 2002.

[32] A. Jurado-Navas, J. M. Garrido-Balsells, J. Francisco Paris, M. Castillo-Vázquez, and A. Puerta-Notario, "Impact of pointing errors on the performance of generalized atmospheric optical channels," *Opt. Exp.*, vol. 20, no. 11, pp. 12550–12562, 2012.

[33] Weblink: *http://www.cie.co.at/*, 28.02.2012.

[34] O. Bader and C. Lui, "Laser safety and the eye: Hidden hazards and practical pearls," Tech. Report: American Academy of Dermatology, Lion Laser Skin Center, Vancouver and University of British Columbia, Vancouver, B.C., 1996.

[35] "Safety and laser products- Part 1: Equipment classification and requirements," International Electrotechnical Commission, (IEC-60825-1), Ed. 3, 2007.

[36] M. Toyoshima, T. Jono, T. Yamawaki, K. Nakagawa, and A. Yamamoto, "Assessment of eye hazard associated with an optical downlink in freespace laser communications," *Proc. SPIE, Free Space Laser Comm. Tech. XIII*, vol. 4272, 2001.

[37] G. D. Fletcher, T. R. Hicks, and B. Laurent, "The SILEX optical interorbit link experiment," *IEEE J. Elec. & Comm. Engg.*, vol. 3, no. 6, pp. 273–279, 2002.

[38] T. Dreischer, M. Tuechler, T. Weigel, G. Baister, P. Regnier, X. Sembely, and R. Panzeca, "Integrated RF-optical TT & C for a deep space mission," *Acta Astronautica*, vol. 65, no. 11, pp. 1772–1782., 2009.

[39] G. Baister, K. Kudielka, T. Dreischer, and M. Tüchler, "Results from the DOLCE (deep space optical link communications experiment) project," *Proc. SPIE, Free Space Laser Comm. Tech. XXI*, vol. 7199, 2009.

[40] D. E. Smith, M. T. Zuber, H. V. Frey, J. B. Garvin, J. W. Head, and D. O. Muhleman et al., "Mars orbiter laser altimeter: Experiment summary after first year of global mapping of Mars," *J. Geophysic. Research*, vol. 106, no. E10, pp. 23689–23722, 2001.

[41] General Atomics Aeronautical Systems, Inc., *GA-ASI and TESAT Partner to Develop RPA-to-spacecraft Lasercom Link*, 2012.

[42] G. G. Ortiz, S. Lee, S. P. Monacos, M. W. Wright, and A. Biswas, "Design and development of a robust ATP subsystem for the altair UAV-to-ground lasercomm 2.5-Gbps demonstration," *Proc. SPIE, Free Space Laser Comm. Tech. XV*, vol. 4975, 2003.

[43] D. Isbel, F. O'Donnell, M. Hardin, H. Lebo, S. Wolpert, and S. Lendroth, "Mars polar lander/deep space 2," NASA Tech. Report, 1999.

[44] Y. Hu, K. Powell, M. Vaughan, C. Tepte, and C. Weimer et al., "Elevation Information in Tail (EIT) technique for lidar altimetry," *Opt. Exp.*, vol. 15, no. 22, pp. 14504–14515, 2007.

[45] N. Perlot, M. Knapek, D. Giggenbach, J. Horwath, M. Brechtelsbauer, and et al., "Results of the optical downlink experiment KIODO from OICETS satellite to optical ground station oberpfaffenhofen (OGS-OP)," *Proc. SPIE, Free-Space Laser Comm. Tech. XIX and Atmospheric Prop. of Electromag. Waves*, vol. 6457, pp. 645704–1–645704–8, 2007.

[46] R. Beer, T. A. Glavich, and D. M. Rider, "Tropospheric emission spectrometer for the Earth observing system's Aura satellite," *Appl. Opt.*, vol. 40, no. 15, pp. 2356–2367, 2001.

[47] K. E. Wilson and J. R. Lesh, "An overview of galileo optical experiment (GOPEX)," Tech Report: TDA progress Report 42-114, Communication Systems Research Section, NASA, 1993.

[48] K. Nakamaru, K. Kondo, T. Katagi, H. Kitahara, and M. Tanaka, "An overview of Japan's Engineering Test Satellite VI (ETS-VI) project," in *Proc IEEE Communications, Int. Conf. on World Prosperity Through Comm.*, vol. 3, (Boston, MA), pp. 1582 –1586, 1989.

[49] Y. Fujiwaraa, M. Mokunoa, T. Jonoa, T. Yamawakia, K. Araia, M. Toyoshimab, H. Kunimorib, Sodnikc, A. Birdc, and B. Demelenned, "Optical inter-orbit communications engineering test satellite (oicets)," *Acta Astronautica, Elsevier*, vol. 61, no. 1-6, pp. 163–175, 2007.

[50] K. Pribil and J. Flemmig, "Solid state laser communications in space (solacos) high data rate satellite communication system verification program," *Proc. SPIE, Space Instrum. and Spacecraft Optics*, vol. 2210, no. 39, 1994.

[51] Z. Sodnik, H. Lutz, B. Furch, and R. Meyer, "Optical satellite communications in Europe," *Proc. SPIE, Free Space Laser Comm. Tech. XXII*, vol. 7587, 2010.

[52] D. M. Boroson, A. Biswas, and B. L. Edward, "MLCD: Overview of NASA's Mars laser communications demonstration system," *Proc. SPIE, Free Space Laser Comm. Tech. XVI*, vol. 5338, 2004.

[53] M. A. Khalighi and M. Uysal, "Survey on free space optical communication: A communication theory perspective," *IEEE Comm. Surve. & Tut.*, vol. 16, no. 4, pp. 2231–2258, 2014.

[54] S. Bloom, E. Korevaar, J. Schuster, and H. Willebrand, "Understanding the performance of free-space optics," *J. Opt. Netw. (OSA)*, vol. 2, no. 6, pp. 178–200, 2003.

[55] H. Demers, F.and Yanikomeroglu and M. St-Hilaire, "A survey of opportunities for free space optics in next generation cellular networks," *IEEE Proc., Ninth Annual Communication Networks and Services Research Conference*, pp. 210–216, 2011.

[56] H. Weichel, *Laser Beam Propagation in the Atmosphere*. SPIE, Bellingham, WA, 1990.

[57] R. K. Long, “Atmospheric attenuation of ruby lasers,” *Proc. of the IEEE*, vol. 51, no. 5, pp. 859–860, May 1963.

[58] R. M. Langer, “Effects of atmospheric water vapour on near infrared transmission at sea level,” in *Report on Signals Corps Contract DA-36-039-SC-723351,* J.R.M. Bege Co., Arlington, Mass, May 1957.

[59] F. G. Smith, J. S. Accetta, and D. L. Shumaker, “The infrared & electro-optical systems handbook: Atmospheric propagation of radiation,” *SPIE press*, vol. 2, 1993.

[60] F. X. Kneizys, *Atmospheric Transmittance/Radiance [Microform]: Computer Code LOWTRAN 6.* Bedford (USA), 1983.

[61] M. Rouissat, A. R. Borsali, and M. E. Chiak-Bled, “Free space optical channel characterization and modeling with focus on Algeria weather conditions,” *Int. J. Comp. Netw. and Infor. Secur.*, vol. 3, pp. 17–23, 2012.

[62] H. Willebrand and B. S. Ghuman, *Free Space Optics: Enabling Optical Connectivity in Today’s Networks*. Sams Publishing, 2002.

[63] R. N. Mahalati and J. M. Kahn, “Effect of fog on free-space optical links employing imaging receivers,” *Opt. Exp.*, vol. 20, no. 2, pp. 1649–661, 2012.

[64] J. M. Wallace and P. V. Hobbs, *Atmospheric Science: An Introductory Survey*. Academic Press, 1977.

[65] I. I. Kim and M. Achour, “Free-space links address the last-mile problem,” vol. 37, 2001.

[66] I. I. Kim, B. McArthur, and E. Korevaar, “Comparison of laser beam propagation at 785 nm and 850 nm in fog and haze for optical wireless communications,” *Proc. SPIE, Opt. wireless comm. III*, vol. 4214, pp. 26–37, 2000.

[67] M. S. Awan, P. Brandl, E. Leitgeb, F. Nadeem, T. Plank, and C. Capsoni, “Results of an optical wireless ground link experiment in continental fog and dry snow conditions,” in *10th Int. Conf. on Telecom. (CONTEL-2009)*, (Zagreb), pp. 45–49, Jun 2009.

[68] C. C. Chen, “Atteumation of electromagnetic radiation by haze, fog, cloud and rain,” Tech. Report: R-1694-PR, United States of Air Force Project Rand, 1975.

[69] I. I. Kim and E. Korevaar, “Availability of free space optics (FSO) and hybrid FSO/RF systems,” Lightpointe Tech. Report, (Weblink: http://www.opticalaccess.com).

[70] A. Z. Suriza, I. M. Rafiqul, A. K. Wajdi, and A. W. Naji, “Proposed parameters of specific rain attenuation prediction for free space optics link operating in tropical region,” *J. of Atmosp. and Solar-Terres. Phys.*, vol. 94, pp. 93–99, 2013.

[71] A. Vavoulas, H. G. Sandalidis, and D. Varoutas, “Weather effects on FSO network connectivity,” *J. Opt. Comm. and Net.*, vol. 4, no. 10, pp. 734–740, 2012.

[72] R. K. Crane and P. C. Robinson, “ACTS propagation experiment: rain-rate distribution observations and prediction model comparisons,” *Proc. IEEE*, vol. 86, no. 6, pp. 946–958, 1997.

[73] “Prediction methods required for the design fo terrestrial free-space optical links,” Recommendation ITU-R P.1814, 2007.

[74] L. C. Andrews and R. L. Phillips, *Laser Beam Propagation through Random Media*. SPIE Press, 2005.

[75] L. C. Andrews, R. L. Phillips, and C. Y. Hopen, *Laser Beam Scintillation with Applications*. SPIE Press, 2001.

[76] R. Rui-Zhong, “Scintillation index of optical wave propagating in turbulent atmosphere,” *Chinese Phy. B*, vol. 18, no. 2, 2009.

[77] E. S. Oh, J. C. Ricklin, G. C. Gilbreath, N. J. Vallestero, and F. D. Eaton, “Optical turbulence model for laser propagation and imaging applications,” *Proc. SPIE, Free Space Laser Comm. and Active Laser Illumina. III*, vol. 5160, pp. 25–32, 2004.

[78] E. Oh, J. Ricklin, F. Eaton, C. Gilbreath, S. Doss-Hammel, C. Moore, J. Murphy, Y. Han Oh, and M. Stell, “Estimating atmospheric turbulene using the PAMELA model,” *Proc SPIE, Free Space Laser Comm. IV*, vol. 5550, pp. 256–266, 2004.

[79] S. Doss-Hammel, E. Oh, J. Ricklinc, F. Eatond, C. Gilbreath, and D. Tsintikidis, “A comparison of optical turbulence models,” *Proc. SPIE, Free Space Laser Comm. IV*, vol. 5550, pp. 236–246, 2004.

[80] S. Karp, R. M. Gagliardi, S. E. Moran, and L. B. Stotts, *Optical Channels: Fibers, Clouds, Water, and the Atmosphere*. Plenum Press, New York and London, 1988.

[81] R. E. Hufnagel and N. R. Stanley, “Modulation transfer function associated with image transmission through turbulence media,” *J. Opt. Soc. Am.*, vol. 54, no. 52, pp. 52–62, 1964.

[82] R. K. Tyson, “Adaptive optics and ground-to-space laser communication,” *Appl. Opt.*, vol. 35, no. 19, pp. 3640–3646, 1996.

[83] R. E. Hufnagel, “Variations of atmospheric turbulence,” Tech. Report, 1974.

[84] G. C. Valley, “Isoplanatic degradation of tilt correction and short-term imaging systems,” *Appl. Opt.*, vol. 19, no. 4, pp. 574–577, February 1980.

[85] A. S. Gurvich, A. I. Kon, V. L. Mironov, and S. S. Khmelevtsov, *Laser radiation in turbulent atmosphere*. Nauka Press, Moscow, 1976.

[86] M. R. Chatterjee and F. H. A. Mohamed, “Modeling of power spectral density of modified von Karman atmospheric phase turbulence and acousto-optic chaos using scattered intensity profiles over discrete time intervals,” *Proc. SPIE, Laser Comm. and Prop. through the Atmosp. and Oce. III*, vol. 9224, 2014.

[87] V. I. Tatarskii, *The Effects of the Turbulent Atmosphere on Wave Propagation*. 1971.

[88] M. C. Roggermann and B. M. Welsh, *Imaging through turbulence*. CRC Press, Boca Raton, FL., 1996.

[89] A. Majumdar and J. Ricklin, *Free-Space Laser Communications: Principles and Advances,*, vol. 2. Springer, 2008.

[90] H. Hemmati, ed., *Near-Earth Laser Communications*. CRC Press, 2009.

[91] B. Beland, *The Infrared and Electro-Optical System Handbook*, vol. 2. SPIE Press, 1993.

[92] T. E. Van Zandt, K. S. Gage, and J. M. Warnock, “An improve model for the calculation of profiles of wind, temperature and humidity,” in *Twentieth Conf. on Radar Meteoro., American Meteoro. Soc., Boston MA*, pp. 129–135, 1981.

[93] E. M. Dewan, G. R. E., B. Beland, and J. Brown, “A model for $C_n^2$ (optical turbulence) profiles using radiosonde data,” Environmental Research Paper-PL-TR-93-2043 1121, Phillips Laboratory, Hanscom, Airforce base, MA, 1993.

[94] J. Parikh and V. K. Jain, “Study on statistical models of atmospheric channel for FSO communication link,” in *Nirma University Int. Conf. on Eng.*-(NUiCONE), pp. 1–7, 2011.

[95] H. G. Sandalidis, “Performance analysis of a laser ground-station-to-satellite link with modulated gamma-distributed irradiance fluctuations,” *J. Opt. Comm. and Net.*, vol. 2, no. 11, pp. 938–943, November 2010.

[96] J. Park, E. Lee, and G. Yoon, “Average bit-error rate of the Alamouti scheme in gamma-gamma fading channels,” *IEEE Photon. Tech. Lett.*, vol. 23, no. 4, pp. 269–271, February 2011.

[97] N. D. Chatzidiamantis, H. G. Sandalidis, G. Karagiannidis, and S. A. Kotsopoulos, “New results on turbulence modeling for free-space optical systems,” *Proc IEEE, Int. Conf. on Tele. Comm.*, pp. 487–492, 2010.

[98] M. A. Kashani, M. Uysal, and M. Kavehrad, *A Novel Statistical Channel Model for Turbulence-Induced Fading in Free-Space Optical Systems*. PhD thesis, Cornell University, 2015.

[99] H. Kaushal, V. Kumar, A. Dutta, H. Aennam, H. Aennam, V. Jain, S. Kar, and J. Joseph, “Experimental study on beam wander under varying atmospheric turbulence conditions,” *IEEE Photon. Tech. Lett.*, vol. 23, no. 22, pp. 1691–1693, 2011.

[100] G. Thuillier, M. Herse, D. Labs, T. Foujols, W. Peetermans, D. Gillotay, P. C. Simon, and H. Mandel, *The solar spectral irradiance from 200 to 2400 nm as measured by the SOLSPEC spectrometer from the ATLAS and EURECA missions*, vol. 214. Kluwer Academic Publisher, Netherlands, 2003.

[101] X. Liu, “Free-space optics optimization models for building sway and atmospheric interference using variable wavelength,” *IEEE Trans. Comm.*, vol. 57, no. 2, pp. 492–498, 2009.

[102] A. Katsuyoshi, “Overview of the optical inter-orbit communications engineering test satellite (OICETS) project,” *J. Nat. Inst. of Info. and Comm. Tech.*, vol. 59, pp. 5–12, 2012.

[103] T. T. Nielsen and G. Oppenhauser, “In-orbit test result of an operational optical intersatellite link between ARTEMIS and SPOT4, SILEX,” *Proc. SPIE, Free Space Laser Comm. Tech. XIV*, vol. 4635, Apr. 2002.

[104] J. Horwath, F. David, M. Knapek, and N. Perlot, “Comparison of link selection algorithms for FSO/RF hybrid network,” *Proc. SPIE, Free Space Laser Comm. Tech. XVII*, vol. 5712, pp. 25–26, 2005.

[105] V. Sannibale and W. H. Ortiz, G.and Farr, “A sub-hertz vibration isolation platform for a deep space optical communication transceiver,” *Proc. SPIE, Free Space Laser Comm. Tech. XXI*, vol. 7199, 2009.

[106] J. Horwath and C. Fuchs, “Aircraft to ground unidirectional laser-communications terminal for high-resolution sensors,” *Proc. SPIE , Free-Space Laser Comm. Tech. XXI*, vol. 7199, 2009.

[107] P. W. Scott and P. W. Young, “Impact of temporal fluctuations of signal-to-noise ratio (burst error) on free-space laser communication

system design," *Proc. SPIE, Opt. Tech. for Comm. Sat. Applic.*, vol. 616, pp. 174–181, 1986.
[108] S. Arnon, "Performance limitations of free-space optical communication satellite networks due to vibrations - Analog case," *Opt. Engg.*, vol. 36, no. 1, 1997.
[109] H. Guo, B. Luo, Y. Ren, S. Zhao, and A. Dang, "Influence of beam wander on uplink of ground-to-satellite laser communication and optimization for transmitter beam radius.," *Opt. Lett.*, vol. 35, no. 12, pp. 1977–1979, 2010.
[110] G. A. Tyler, "Bandwidth considerations for tracking through turbulence," *J. Opt. Soc. Am.*, vol. 11, no. 1, pp. 358–367, 1994.
[111] H. Hemmati, "Interplanetary laser communications," in *Optics & Photonics News (OSA)*, 2007.
[112] H. J. Kramer, *Observation of the Earth and Its Environment: Survey of Missions and Sensors*. Springer, 2002.
[113] N. S. Kopeika, A. Zilberman, and Y. Sorani, "Measured profiles of aerosols and turbulence for elevations of 2-20 km and consequences on widening of laser beams," *Proc. SPIE, Opt. Pulse and Beam Prop. III*, vol. 4271, no. 43, pp. 43–51, January 2001.
[114] A. Zilberman, N. S. Kopeika, and Y. Sorani, "Laser beam widening as a function of elevation in the atmosphere for horizontal propagation," *Proc. SPIE, Laser Weapons Tech. II*, vol. 4376, no. 177, pp. 177–188, April 2001.
[115] V. I. Tatarskii, *Wave propagation in a Turbulent Medium*. McGraw-Hill, New York, 1961.
[116] P. J. Titterton, "Power reduction and fluctuations caused by narrow laser beam motion in the far field," *Appl. Opt.*, vol. 12, no. 2, pp. 423–425, 1973.
[117] L. C. Andrews, R. L. Phillips, and C. Y. Hopen, "Scintillation model for a satellite communication link at large zenith angles," *Opt. Eng.*, vol. 39, pp. 3272–3280, 2000.
[118] J. Katz, "Planets as background noise sources in free space optical communiction," TDA Progress Report: 42-85, Advanced Electronic Material and Devices Section, 1986.
[119] F. Roddier, "The effects of atmospheric turbulence in optical astronomy," in *Progress in Optics XIX, North Holland, New York*, 1981.
[120] J. H. Churnside, "Aperture averaging of optical scintillation in the turbulent atmosphere," *Appl. Opt.*, vol. 30, no. 15, pp. 1982–1994, 1991.
[121] L. C. Andrews, "Aperture averaging of optical scintillations: power fluctuations and temporal spectrum," *Waves in random media*, vol. 10, pp. 53–70, 2000.
[122] L. C. Andrews, "Aperture-averaging factor for optical scintillations of plane and spherical waves in the atmosphere," *J. Opt. Soc. Am.*, vol. 9, no. 4, pp. 597–600, 1992.
[123] N. Perlot and D. Fritzsche, "Aperture-averaging, theory and measurements," *Proc. SPIE, Free-Space Laser Comm. Tech. XVI*, vol. 5338, pp. 233–242, 2004.
[124] R. F. Lutomirski and H. T. Yura, "Aperture-averaging factor of a fluctuating light signal," *J. Opt. Soc. Am.*, vol. 59, no. 9, pp. 1247–1248, Sept.1969.
[125] L. M. Wasiczko and C. C. Davis, "Aperture averaging of optical scintillations in the atmosphere: Experimental results," *Proc. SPIE, Atmospheric Prop. II*, vol. 5793, pp. 197–208, 2005.
[126] H. T. Yura and W. G. McKinley, "Aperture averaging of scintillation for space-to-ground optical communication applications," *Appl. Opt.*, vol. 22, no. 11, pp. 1608–1609, Jun. 1983.
[127] G. L. Bastin, L. C. Andrews, R. L. Phillips, R. A. Nelson, B. A. Ferrell, M. R. Borbath, D. J. Galus, P. G. Chin, W. G. Harris, J. A. Marin, G. L. Burdge, D. Wayne, and R. Pescatore, "Measurements of aperture averaging on bit-error-rate," *Proc. SPIE , Atmosphe. Opt. Model., Measur., and Simul.*, vol. 5891, 2005.
[128] F. Strömqvist Vetelino, C. Young, L. Andrews, and J. Recolons, "Aperture averaging effects on the probability density of irradiance fluctuations in moderate-to-strong turbulence," *Appl. Opt.*, vol. 46, no. 11, pp. 2099–2108, 2007.
[129] S. M. Navidpour, M. Uysal, and M. Kavehrad, "BER performance of free-space optical transmission with spatial diversity," *IEEE Trans. Wireless Comm.*, vol. 6, no. 8, pp. 2813–2819, August 2007.
[130] X. Zhu and J. M. Kahn, "Maximum-likelihood spatial-diversity reception on correlated turbulent free-space optical channels," in *IEEE Conf. Global Comm.*, vol. 2, (San Francisco, California), pp. 1237–1241, 2000.
[131] M. Jeganathan, M. Toyoshima, K. E. Wilson, and J. R. Lesh, "Data analysis result from GOLD experiments," *Proc. SPIE, Free Space Laser Comm. Tech. IX*, vol. 2990, pp. 70–81, 1998.
[132] Z. Chen, S. Yu, T. Wang, G. Wu, S. Wang, and W. Gu, "Channel correlation in aperture receiver diversity systems for free-space optical communication," *J. Optics*, vol. 14, no. 12, 2012.
[133] H. Kaushal, V. K. Jain, and S. Kar, "Ground to satellite optical communication link performance with spatial diversity in weak atmospheric turbulence," *J. Fiber and Integr. Opt.*, vol. 29, no. 4, pp. 315–340, 2010.
[134] G. Yang, M. Ali Khalighi, Z. Ghassemlooy, and S. Bourennane, "Performance evaluation of receive-diversity free-space optical communications over correlated Gamma-Gamma fading channels," *Appl. Opt.*, vol. 52, no. 24, pp. 5903–5911, 2013.
[135] M. K. Simon and V. A. Vilnrotter, "Alamouti-type space time coding for free space optical communication with direct detection," *IEEE Trans. Comm.*, vol. 50, no. 8, pp. 1293–1300, 2002.
[136] Z. Hajjarian and J. Fadlullah, "MIMO free space optical communications in turbid and turbulent atmosphere," *J. Comm.*, vol. 4, no. 8, pp. 524–532, 2009.
[137] V. Vilnrotter and M. Srinivasan, "Adaptive detector arrays for optical communications receivers," *IEEE Trans. Comm.*, vol. 50, no. 7, pp. 1091–1097, 2002.
[138] N. Letzepis and A. Guilléni Fàbregas, "Outage probability of the Gaussian MIMO free-space optical channel with PPM," *IEEE Trans. Comm.*, vol. 57, no. 12, pp. 3682–3690, 2009.
[139] N. Letzepis, I. Holland, and W. Cowley, "The Gaussian free space optical MIMO channel with $Q$-ary pulse position modulation," *IEEE Trans. Wireless Comm.*, vol. 7, no. 5, pp. 1744–1753, 2008.
[140] N. Cvijetic, S. G. Wilson, and M. Brandt-Pearce, "Receiver optimization in turbulent free-space optical MIMO channels with APDs and $Q$-ary PPM," *IEEE Photon. Tech. Lett.*, vol. 19, no. 2, pp. 103–105, 2007.
[141] K. Chakraborty, "Capacity of the MIMO optical fading channel," *Proc. IEEE, Int. Symp. Inf. Theory*, pp. 530–534, 2005.
[142] H. Park and J. R. Barry, "Trellis-coded multiple-pulse-position modulation for wireless infrared communications," *IEEE Trans. Comm.*, vol. 52, no. 4, pp. 643–651, 2004.
[143] E. Bayaki, R. Schober, and R. Mallik, "Performance analysis of MIMO free-space optical systems in gamma-gamma fading," *IEEE Trans. Comm.*, vol. 57, no. 11, pp. 3415–3424, 2009.
[144] I. B. Djordjevic, B. Vasic, and M. A. Neifeld, "Multilevel coding in free-space optical MIMO transmission with $Q$-ary PPM over the atmospheric turbulence channel," *IEEE Photon. Tech. Lett.*, vol. 18, no. 14, pp. 1491–1493, 2006.
[145] J. Laneman, D. Tse, and G. W. Wornell, "Cooperative diversity in wireless networks: Efficient protocols and outage behavior," *IEEE Trans. Inf. Theory*, vol. 50, no. 12, pp. 3062–3080, 2004.
[146] M. Karimi and M. Nasiri-kenari, "BER analysis of cooperative systems in free-space optical networks," *J. Lightwave Tech.*, vol. 27, no. 24, pp. 5639–5647, 2009.
[147] M. Safari and M. Uysal, "Relay-assisted free-space optical communication," *IEEE Trans. Wireless Comm.*, vol. 7, no. 12, pp. 5441–5449, 2008.
[148] F. Xu, M. Ali Khalighi, P. Causse, and S. Bourennane, "Performance of coded time-diversity free-space optical links," *IEEE 24th Biennial Symp. on Comm.*, pp. 146–149, 2008.
[149] F. E. Zocchi, "A simple analytical model of adaptive optics for direct detection free-space optical communication," *Opt. Comm.*, vol. 248, no. 4-6, pp. 359–374, April 2005.
[150] P. Barbier, D. Rush, M. Plett, and P. Polak-Dingels, "Performance improvement of a laser communication link incorporating adaptive optics," *Proc. SPIE, The Inter. Society for Opt. Eng.*, vol. 3432, pp. 93–102, 1998.
[151] R. K. Tyson, *Principles of adaptive optics*. Academic, Boston, MA, 1998.
[152] K. E. Wilson, L. P. R., R. Cleis, J. Spinhirne, and R. Q. Fugate, "Results of the compensated Earth-Moon-Earth Retroreflector laser link (CEMERLL) experiments," The Telecommunication and Data Acquisition Progress Report: 42-131, Jet Propulsion Laboratory, Pasadena, California, 1997.
[153] N. B. Baranova, A. V. Mamaev, N. F. Pilipetsky, V. V. Shkunov, and B. Y. Zel'dovich, "Wavefront dislocation: Topological limitations for adaptive systems with phase conjugations," *J. Opt. Soc. Am.*, vol. 73, pp. 525–528, 1983.
[154] D. L. Fried, "Branch point problem in adaptive optics," *J. Opt. Soc. Am.*, vol. 15, pp. 2759–2768, 1998.
[155] M. A. Vorontsov, G. W. Carhart, and J. C. Ricklin, "Adaptive phase distortion correction based on parallel gradient-descent optimization," *Opt. Lett.*, vol. 22, pp. 907–909, 1997.

[156] M. A. Vorontsov and V. P. Sivokon, "Stochastic parallel gradient descent technique for high resolution wave-front phase-distortion correction," *J. Opt. Soc. Am.*, vol. 15, pp. 2745–2758, 1998.

[157] C. A. Thompson, M. W. Kartz, L. M. Flath, S. C. Wilks, R. A. Young, G. W. Johnson, and A. J. Ruggiero, "Free space optical communications utilizing MEMS adaptive optics correction," *Proc. SPIE, Society of Photo-Instru. Eng.*, vol. 4821, 2002.

[158] T. Weyrauch, M. A. Vorontsov, T. G. Bifano, J. A. Hammer, M. Cohen, and G. Cauwenberghs, "Microscale adaptive optics:wave-front control with a $\mu$-mirror array and a VLSI stochastic gradient descent controller," *Appl. Opt.*, vol. 40, pp. 4243–4253, 2001.

[159] D. P. Greenwood, "Bandwidth specifications for adaptive optics systems," *J. Opt. Soc. Am.*, vol. 67, pp. 174–176, 1976.

[160] W. C. Brown, "Optimum thresholds for optical On-Off keying receivers operating in the turbulent atmosphere," *Proc. SPIE, Free Space Laser Comm. Tech. IX*, vol. 2290, pp. 254–261, 1997.

[161] W. E. Webb and J. T. Jr Marino, "Threshold detection in an on-off binary communication channel with atmospheric scintillation," *Appl. Opt.*, vol. 14, pp. 1413–1417, 1975.

[162] X. Zhu and J. M. Kahn, "Pilot-symbol assisted modulation for correlated turbulent free-space optical channels," *Proc. SPIE, Intl. Symp. Optical Science Technol, San Diego, CA*, 2001.

[163] N. D. Chatzidiamantis, G. K. Karagiannidis, and M. Uysal, "Generalized maximum-likelihood sequence detection for photon-counting free space optical systems," *IEEE Trans. Comm.*, vol. 58, no. 10, pp. 3381–3385, 2010.

[164] M. L. B. Riediger, R. Schober, and L. Lampe, "Blind detection of on-off keying for free-space optical communications," *IEEE Canadian Conf. on Electri. and Comp. Eng . CCECE* , pp. 1361–1364, 2008.

[165] N. D. Chatzidiamantis, M. Uysal, T. A. Tsiftsis, and G. K. Karagiannidis, "Iterative near maximum-likelihood sequence detection for MIMO optical wireless systems," *J. Lightwave Tech.*, vol. 28, no. 7, pp. 1064–1070, 2010.

[166] H. Hemmati, *Deep Space Optical Communication*. John Wiley & Sons, New York, 2006.

[167] R. M. Gagliardi and S. Karp, *Optical Communications*. John Wiley & Sons, New York, 1976.

[168] S. Hranilovic and D. A. Johns, "A multilevel modulation scheme for high-speed wireless infrared communications," *Proc. IEEE Int. Symp. Cir. & Sysm.*, vol. 6, pp. 338–341, 1999.

[169] R. M. Gagliardi and S. Karp, "*M*-ary Poisson detection and optical communications," *IEEE Trans. Comm.*, vol. 17, no. 2, pp. 208–216, 1969.

[170] B. Reiffen and H. Sherman, "An optimum demodulator for Poisson processes: Photon source detectors," *Proc. IEEE*, vol. 51, pp. 1316–1320, 1963.

[171] A. Aladeloba, A. Phillips, and M. S. Woolfson, "DPPM FSO communication systems impaired by turbulence, pointing error and ASE noise," (Coventry), pp. 1–4, 2012.

[172] D. Zwillinger, "Differential PPM has a higher throughput than PPM for the bandlimited and average power limited optical channels," *IEEE Trans. Inf. Theory*, vol. 34, pp. 1269–1273, 1988.

[173] P. Gopal, V. K. Jain, and S. Kar, "Performance analysis of ground to satellite FSO system with DAPPM scheme in weak atmospheric turbulence," in *Int. Conf. on Fiber Optics and Photon. (OSA)*, 2012.

[174] C. Liu, Y. Yao, J. Tian, Y. Yuan, Y. Zhao, and B. Yu, "Packet error rate analysis of DPIM for free-space optical links with turbulence and pointing errors," *Chinese Opt. Lett.*, vol. 12, pp. S10101–11–5, 2014.

[175] M. Faridzadeh, A. Gholami, Z. Ghassemlooy, and S. Rajbhandari, "Hybrid PPM-BPSK subcarrier intensity modulation for free space optical communications," in *Proc. IEEE, 16th European Conf. on Netw. and Opt. Comm.*, (Newcastle-Upon-Tyne), pp. 36–39, 2011.

[176] Z. Ghassemlooy, W. O. Popoola, V. Ahmadi, and E. Leitgeb, "MIMO free-space optical communication employing subcarrier intensity modulation in atmospheric turbulence channels," *Lec. notes of the Inst. for Comp. Scien., Social Inform. and Telecomm. Eng.- Invited talk*, vol. 16, pp. 61–73, 2009.

[177] J. H. Sinsky, A. Adamiecki, A. Gnauck, C. Jr. Burrus, J. Leuthold, O. Wohlgemuth, and A. Umbach, "RZ-DPSK transmission using a 42.7-Gb/s integrated balanced optical front end with record sensitivity," *J. Lightwave Tech.*, vol. 22, no. 1, 2003.

[178] W. A. Atia and R. S. Boundurant, "Demonstration of return-to-zero signaling in both OOK and DPSK formats to improve receiver sensistivity in an optically preamplified receiver," *LEOS*, 1999.

[179] A. Gnauck, G. Raybon, S. Chandrasekhar, J. Leuthold, C. Doerr, L. Stulz, A. Agarwal, S. Banerjee, D. Grosz, S. Hunsche, A. Kung, A. Marhelyuk, D. Maywar, M. Movassaghi, X. Liu, C. Xu, X. Wei, and D. Gill, "2.5 Tb/s (64 x 42.7 Gb/s) transmission over 40x100 km NZDSF using RZ-DPSK format and all-Raman-amplified spans," *Proc. IEEE. Opt. Fiber Comm. Conf. and Exh. (OFC)*, 2002.

[180] J. Hagenauer and E. Lutz, "Forward error correction coding for fading compensation in mobile satellite channels," *IEEE J. Sel. Areas in Comm.*, vol. 5, no. 2, pp. 215–225, 1987.

[181] J. A. Anguita, I. B. Djordjevic, M. A. Neifeld, and B. V. Vasic, "High-rate error-correction codes for the optical atmospheric channel," *Proc. SPIE, Free Space Laser Comm. Tech. V*, vol. 5892, 2005.

[182] J. A. Alzubi, O. A. Alzubi, and T. M. Chen, *Forward Error Correction Based on Algebraic Geometric Theory*. Springer, 2012.

[183] F. Xu, M. A. Khalighi, P. Causse, and S. Bourennane, "Performance of coded time-diversity free space optical links," *Proc IEEE, Biennial Symposium on Communications*, pp. 146–149, 2008.

[184] J. Nakai, *Coding and modulation analysis for optical communication channels*. PhD thesis, MIT Cambridge, MA, 1982.

[185] F. M. Davidson and Y. T. Koh, "Interleaved convolutional coding for the turbulent atmospheric optical communication channel," *IEEE Trans. Comm.*, vol. 36, no. 9, pp. 993–1003, 1988.

[186] J. A. Anguita, I. B. Djordjevic, M. A. Neifeld, and B. V. Vasic, "Shannon capacities and error-correction codes for optical atmospheric turbulent channels," *J. Opt. Netw. (OSA)*, vol. 4, pp. 586–601, 2005.

[187] S. S. Muhammad, T. javornik, I. Jelovcan, E. Leitgeb, and O. Koudelka, "Reed solomon coded PPM for terrestrial FSO links," in *Proc. Int. Conf. on Electr. Eng. (ICEE)*, 2007.

[188] C. Fewer, M. Flanagan, and A. Fagan, "A versatile variable rate LDPC codec architecture," *IEEE Trans. on Ckts and Sys.*, vol. 54, no. 10, pp. 2240–2251, 2007.

[189] N. Minseok, N. Kim, P. Hyuncheol, and L. Hyuckjae, "A variable rate LDPC coded V-BLAST system," *Proc IEEE, Vehicu. Tech. Conf. VTC*, vol. 4, pp. 2540–2543, 2004.

[190] I. B. Djordjevic, B. Vasic, and M. A. Niefeld, "Power efficient LDPC coded modulation for free-space optical communication over the atmospheric turbulence channel," *Proc IEEE, Optical Fiber Communication and the National Fiber Optic Engineers*, pp. 1–3, 2007.

[191] B. Barua and S. P. Majumder, "LDPC coded FSO communication system under strong turbulent condition," *Proc IEEE, Int. Conf. Electri. & Comp. Eng. (ICECE)*, pp. 414–417, 2012.

[192] I. B. Djordjevic, "LDPC-coded MIMO optical communication over the atmospheric turbulence channel using $Q$-ary pulse position modulation," *Opt. Exp.*, vol. 16, pp. 10026–10032, 2007.

[193] B. Barua and D. Barua, "Analysis the performance of a LDPC coded FSO system with $Q$-ary pulse-position modulation," in *Proc. IEEE, 3rd Int. Conf. on Comp. Resear. and Develop.(ICCRD)*, vol. 1, (Shanghai), pp. 339–343, 2011.

[194] I. B. Djordjevic, B. Vasic, and M. A. Neifeld, "LDPC coded OFDM over the atmospheric turbulence channel," *Opt. Exp.*, vol. 15, no. 10, pp. 6332–6346, 2007.

[195] S. Benedetto, G. Montorsi, D. Divsalr, and F. Pollara, "Soft-output decoding algorithm in iterative decoding of Turbo code," TDA Progress Report: 42-124, 1996.

[196] J. Hagenauer, "Source-controlled channel decoding," *IEEE Trans. Comm.*, vol. 43, no. 9, pp. 2449–2457, 1995.

[197] H. R. Sadjadpour, "Maximum a posteriori decoding algorithms for Turbo codes," *Proc SPIE, Digital Wireless Comm. II*, vol. 4045, pp. 73–83, 2000.

[198] L. Yang, J. Cheng, and J. F. Holzman, "Performance of convolutional coded OOK IM/DD system over strong turbnulence channels," in *Int. Conf. on Compu. Netw. and Comm.*, 2013.

[199] H. Moradi, M. Falahpour, H. H. Refai, and P. G. LoPresti, "BER analysis of optical wireless signals through lognormal fading channels with perfect CSI," in *Proc. IEEE, Int. Conf. Telecomm. (ICT)*, pp. 493–497, 2010.

[200] X. Zhu and J. M. Kahn, "Free-space optical communication through atmospheric turbulence channels," *IEEE Trans. Comm.*, vol. 50, no. 8, pp. 1293–1300, 2002.

[201] N. Kumar, V. Jain, and S. Kar, "Evaluation of the performance of FSO system using OOK and M-PPM modulation schemes in inter-satellite links with Turbo codes," in *Proc. IEEE, Int. Conf. Electr., Commu. and Comp. Techn.*, pp. 59–63, 2011.

[202] H. G. Sandalidis, T. A. Tsiftsis, G. K. Karagiannidis, and M. Uysal, "BER performance of FSO links over strong atmospheric turbulence channels with pointing errors," *IEEE Comm. Lett.*, vol. 12, no. 1, pp. 44–46, 2008.

[203] M. Uysal and L. Jing, "Error performance analysis of coded wireless optical links over atmospheric turbulence channels," in *Proc. IEEE Wireless Comm. and Netw.*, vol. 4, pp. 2405–2410, 2004.

[204] W. Zixiong, W. D. Zhong, and Y. Changyuan, "Performance improvement of OOK free-space optical communication systems by coherent detection and dynamic decision threshold in atmospheric turbulence conditions," *IEEE Photon. Tech. Lett.*, vol. 24, no. 22, pp. 2035–2037, 2012.

[205] M. Abaza, N. A. Mohammed, and M. H. Aly, "BER performance of *M*-ary PPM free-space optical communications with channel fading," in *High Capacity Optical Networks and Enabling Technologies (HONET)*, (Riyadh), pp. 111–115, 2011.

[206] Y. Xiang, L. Zengji, Y. Peng, and S. Tao, "BER performance analysis for *M*-ary PPM over Gamma-Gamma atmospheric turbulence channels," in *Proc. IEEE, Wireless Comm.Netw. and Mobile Comp. (WiCOM)*, (Chengdu), 2010.

[207] G. Z. Antonio, C. V. Carmen, and C. V. Beatriz, "Space-time trellis coding with transmit laser selection for FSO links over strong atmospheric turbulence channels," *Opt. Exp.*, vol. 18, no. 6, pp. 5356–5366, 2010.

[208] T. T. Nguyen and L. Lampe, "Coded multipulse pulse-position modulation for free-space optical communications," *IEEE Trans. Comm.*, vol. 20, no. 20, pp. 1–6, 2009.

[209] J. Li, A. Hylton, J. Budinger, J. Nappier, J. Downey, and D. Raible, "Dual-pulse pulse position modulation (DPPM) for deep-space optical communications: Performance and practicality analysis," Tech. Report: NASA/TM-2012-216042, National Aeronautics and Space Administration, Ohio, 2012.

[210] A. Viswanath, H. Kaushal, V. K. Jain, and S. Kar, "Evaluation of performance of ground to satellite free space optical link under turbulence conditions for different intensity modulation schemes," *Proc. SPIE, Free Space Laser Comm. and Atmosph. Prop. (XXVI)*, vol. 8971, 2014.

[211] W. Gappmair and S. Muhammad, "Error performance of terrestrial FSO links modelled as PPM/Poisson channels in turbulent atmosphere," *IEEE Electr. Lett.*, vol. 43, no. 5, pp. 63–64, 2007.

[212] M. R. Bhatnagar, "Differential decoding of SIM DPSK over FSO MIMO links," *IEEE Comm. Lett.*, vol. 17, no. 1, pp. 79–82, 2013.

[213] K. Prabu, S. Bose, and D. S. Kumar, "BPSK based subcarrier intensity modulated free space optical system in combined strong atmospheric turbulence," *Optics Comm. (Elsevier)*, vol. 305, pp. 185–189, 2013.

[214] X. Zhu and J. M. Kahn, "Performance bounds for coded free-space optical communications through atmospheric turbulence channels," *IEEE Trans. Comm.*, vol. 51, no. 8, pp. 1233–1239, 2003.

[215] M. Z. Hassan, T. A. Bhuiyan, S. M. Shahrear Tanzil, and S. P. Majumder, "Turbo-coded MC-CDMA communication link over strong turbulence fading limited FSO channel with receiver space diversity," *J. ISRN Comm. and Net.*, vol. 26, 2011.

[216] F. Xu, M. A. Khalighi, and S. Bourennane, "Coded PPM and multipulse PPM and iterative detection for free-space optical links," *J. Opt. Comm. and Net.*, vol. 1, no. 5, pp. 404–415, 2009.

[217] H. Sandhu and D. Chadha, "Terrestrial free space LDPC coded MIMO optical link," in *Proc. World Cong. on Eng. and Comp. Sci. (WCECS)*, vol. 1, 2009.

[218] H. Sandhu and D. Chadha, "Power and spectral efficient free space optical link based on MIMO system," in *Proc. IEEE Annual Comm. Net. and Serv. Resea (CNSR'08)*, pp. 504–509, 2008.

[219] H. Ai-ping, F. Yang-Yu, L. Yuan-Kui, J. Meng, B. Bo, and T. Qing-Gui, "A differential pulse position width modulation for optical wireless communication," in *Proc. IEEE Indus. Electroni and App.*, (Xi'an), pp. 1773–1776, 2009.

[220] E. Ali, V. Sharma, and P. Hossein, "Hybrid channel codes for efficient FSO/RF communication systems," *IEEE. Trans. Comm.*, vol. 58, no. 10, pp. 2926–2938, 2010.

[221] Y. T. Koh and F. Davidson, "Interleaved concatenated coding for the turbulent atmospheric direct detection optical communication channel," *IEEE Trans. Comm.*, vol. 37, no. 6, pp. 648–651, 2002.

[222] N. T. Dang and A. T. Pham, "Performance improvement of FSO/CDMA systems over dispersive turbulence channel using multi-wavelength PPM signaling," *Opt. Exp.*, vol. 20, no. 24, 2012.

[223] G. G. Ortiz, J. V. Sandusky, and A. Biswas, "Design of an opto-electronic receiver for deep-space optical communications," Tech. Report: TMO Progress Report 42-142, 2000.

[224] G. A. Mahdiraji and E. Zahedi, "Comparison of selected digital modulation schemes (OOK, PPM and DPIM) for wireless optical communications," *Proc IEEE, 4th Student Conf. on Resear. and Develop. (SCOReD),*, pp. 5–10, 2006.

[225] S. Lee, K. E. Wilson, and M. Troy, "Background noise mitigation in deep-space optical communications using adaptive optics," Tech. Report: IPN progress report 42-161, Jet Propulsion Laboratory, California Institute of Technology, 2005.

[226] A. J. Hashmi, A. A. Eftekhar, A. Adibi, and F. Amoozegar, "Analysis of adaptive optics-based telescope arrays in a deep-space inter-planetary optical communications link between Earth and Mars," *Optics Comm. (Elsevier)*, vol. 333, pp. 120–128, 2014.

[227] K. Wilson, M. Troy, V. Vilnrotter, M. Srinivasan, B. Platt, M. Wright, V. Garkanian, and H. Hemmati, "Daytime adaptive optics for deep space optical communication," in *Space activities and cooperation contributing to all Pacific basin countries-ISCOPS*, (Japan), 2003.

[228] W. Haiping and M. Kavehrad, "Availability evaluation of ground-to-air hybrid FSO/RF links," *J. Wireless Information Networks (Springer)*, vol. 14, no. 1, pp. 33–45, 2007.

[229] H. Moradi, M. Falahpour, H. H. Refai, P. G. LoPresti, and M. Atiquzzaman, "On the capacity of hybrid FSO/RF links," *Proc IEEE, Globecom*, 2010.

[230] Y. Tang, M. Brandt-Pearce, and S. Wilson, "Adaptive coding and modulation for hybrid FSO/RF systems," in *Proc. IEEE, 43rd Asilomar Conf. on Sig., Sys. and Comp.*, (Pacific Grove, CA), 2009.

[231] D. K. Kumar, Y. S. S. R. Murthy, and G. V. Rao, "Hybrid cluster based routing protocol for free-space optical mobile ad hoc networks (FSO/RF MANET)," *Proc. Int. Conf. of Intell. Comput. (Springer-Verlag)*, vol. 199, pp. 613–620, 2013.

[232] J. Derenick, C. Thorne, and J. Spletzer, "Hybrid free-space optics/radio frequency (FSO/RF) networks for mobile robot teams," Tech. Report, Lehigh University Bethlehem, PA USA.

[233] S. Chia, M. Gasparroni, and P. Brick, "The next challenge for cellular networks: backhaul," *Proc IEEE, Microwave Mag.*, vol. 10, no. 5, pp. 54–66, 2009.

[234] C. Milner, S.D.and Davis, "Hybrid free space optical/RF networks for tactical operations," in *Military Comm. Conf. (MILCOM)*, 2004.

[235] A. Kashyap and M. Shayman, "Routing and traffic engineering in hybrid RF/FSO networks," in *IEEE Int. Conf. on Comm.*, 2005.

[236] B. Liu, Z. Liu, and D. Towsley, "On the capacity of hybrid wireless network," in *IEEE INFOCOM'03*, 2003.

[237] Patent: US 2014/0233960 A1, ed., *Dynamic packet redundancy for a free space optical communication link*, 2014.

[238] V. V. Mai, T. C. Thang, and A. T. Pham, "Performance of TCP over free-space optical atmospheric turbulence channels," *J. Opt. Comm. and Net.*, vol. 5, no. 11, pp. 1168–1177, 2013.

[239] S. Tati and T. F. L. Porta, "Design of link layer protocol for free space optical links," in *Pennsylvania State University*.

[240] H. Moradi, M. Falahpour, H. H. Refai, P. G. Lopresti, and M. Atiquzzaman, "Availability modeling of FSO/RF mesh networks through turbulence-induced fading channels," *IEEE Comm. Soc.*, 2010.

[241] K. Kiasaleh, "Hybrid ARQ for FSO communications through turbulent atmosphere," *IEEE Comm. Lett.*, vol. 14, no. 9, pp. 866–868, 2010.

[242] I. F. Akyildiz and W. Liu, "A general analysis technique for ARQ protocol performance in high speed networks," *Proc. IEEE*, pp. 498–507, 1999.

[243] O. Gürbüz and O. Tuzla, "A transparent ARQ scheme for broadcast wireless access," *Proc IEEE, Wireless Comm. and Netw. Conf. (WCNC)*, vol. 1, pp. 423 –429, 2004.

[244] K. S. Pathapati, J. P. Rohrer, and J. P. G. Sterbenz, "End-to-End ARQ: Transport-layer reliability for airborne telemetry networks," in *Inter. Telemeter. Conf. (ITC)* , 2010.

[245] A. Hammons and F. Davidson, "On the design of automatic repeat request protocols for turbulent free-space optical links," in *Military Comm. Conf. (MILCOM)*, pp. 808–813, 2010.

[246] E. J. Lee and V. Chan, "Performance of the transport layer protocol for diversity communication over the clear turbulent atmospheric optical channel," in *Proc IEEE, Int. Conf. on Comm.-(ICC)*, vol. 1, pp. 333–339, 2005.

[247] V. M. Vuong, C. T. Truong, and T. P. Anh, "Performance analysis of TCP over free-space optical links with ARQ-SR," in *18th European Conf. on Netw. and Opt. Comm. & 8th Conf. on Opt. Cabling and Infrastruc.*, pp. 105–112, 2013.

[248] C. Kose and T. R. Halford, "Incremental redundancy hybrid ARQ protocol design for FSO links," *Proc. IEEE, Military Comm. Conf. (MILCOM)*, pp. 1–7, 2009.

[249] E. Zedini, A. Chelli, and M. S. Alouini, "On the performance analysis of hybrid ARQ with incremental redundancy and with code combining over free-space optical channels with pointing errors," *IEEE J. Photon.*, vol. 6, no. 4, 2014.

[250] G. Caire and D. Tuninetti, "The throughput of hybrid-ARQ protocols for the Gaussian collision channel," *IEEE Trans. Inf. Theory*, vol. 47, no. 5, pp. 1971–1988, 2001.

[251] S. Aghajanzadeh and M. Uysal, "Information theoretic analysis of hybrid-ARQ protocols in coherent free-space optical system," *IEEE. Trans. Comm.*, vol. 60, no. 5, pp. 1432–1442, 2012.

[252] C. A. Rjeily and S. Haddad, "Cooperative FSO systems: Performance analysis and optimal power allocation," *J. Lightwave Tech.*, vol. 29, no. 7, pp. 1058–1065, 2011.

[253] V. V. Mai and A. T. Pham, "Performance analysis of cooperative-ARQ schemes in free-space optical communications," *IEICE Trans. Comm.*, vol. E94-B, no. 1, pp. 1–10, 2011.

[254] R. A. Hammons and F. Davidson, "Diversity rateless round robin for networked FSO communications," *OSA Tech. Digest*, 2011.

[255] A. R. Hammons Jr., "Systems and methods for a Rateless Round Robin Protocol for adaptive error control," US patent: US 20100281331 A1, 2010.

[256] T. A. Courtade and R. D. Wesel, "A cross-layer perspective on rateless coding for wireless channels," *Proc IEEE, ICC*, 2009.

[257] A. R. Hammons Jr. and F. Davidson, "On the design of automatic repeat request protocols for turbulent free-space optical links," in *Military Comm. Conf. (MILCOM)*, 2010.

[258] A. Desai and S. Milner, "Autonomous reconfiguration in free-space optical sensor networks," *IEEE J. Sel. Areas in Comm.*, vol. 23, no. 8, pp. 1556–1563, 2005.

[259] C. C. Davis, I. I. Smolyaninov, and S. D. Milner, "Flexible optical high data rate wireless links and networks," *IEEE Comm. Mag.*, vol. 41, no. 3, pp. 51–57, 2003.

[260] M. Hassan, F. Maryam, H. R. Hazem, G. L. Peter, and M. Atiquzzaman, "Reconfiguration modeling of reconfigurable hybrid FSO/RF links," in *IEEE Int. Conf. on Comm.*, (Cape Town, South Africa), pp. 1–5, 2010.

[261] A. Desai, J. Llorca, and S. D. Milner, "Autonomous reconfiguration of backbones in free-space optical networks," in *Military Comm. Conf. (MILCOM)*, 2004.

[262] J. Llorca, A. Desai, U. Vishkin, C. Davis, and S. Milner, "Reconfigurable optical wireless sensor networks," *Proc. SPIE,Optics in Atmosphe. Prop. and Adaptive Sys. VI*, vol. 5237, 2004.

[263] H. Narra, Y. Cheng, E. K. Çetinkaya, J. P. Rohrer, and J. P. Sterbenz, "Destination-sequenced distance vector (DSDV) routing protocol implementation in NS-3," in *Wireless Snesor Network-3, Barcelona, Spain*, 2011.

[264] T. Clausen and P. Jacque, "Optimized link state routing protocol," RFC 3626, 2003.

[265] S. Murthy and J. J. Garcia-Luna-Aceves, "An efficient routing protocol for wireless networks," *Mobile Networks and Applications (Springer)*, vol. 1, no. 2, pp. 183–197, 1996.

[266] C. E. Perkins and E. M. Royer, "Ad-hoc on-demand distance vector routing," in *Proc IEEE, Workshop on Mobile Comp. Sys. and Appl., (WMCSA)*, pp. 90–100, 1999.

[267] L. Blazevic, J. Le Boudec, and S. Giordano, "A location based routing method for irregular mobile ad hoc networks," EPFL-IC Report Number IC/2003/30, 2003.

[268] N. A. and B. Berk Ustundag, "Improvement of DSR protocol using group broadcasting," *Int J. Comp. Sci. Netw. Secur.*, vol. 10, no. 6, 2010.

[269] M. S. Aminian, *Routing in Terrestrial Free Space Optical Ad- Hoc Networks*. PhD thesis, Linköping University Institute of Technology, Norrköping, 2014.

[270] Y. Jahir, M. Atiquzzaman, H. Refai, and P. G. LoPresti, "AODVH: Ad hoc on-demand distance vector routing for hybrid nodes," Report, University of Oklahoma, Norman, OK and The University of Tulsa, Tulsa, OK.

[271] J. Kulik, W. R. Heinzelman, and H. Balakrishnan, "Negotiation-based protocols for disseminating information in wireless sensor networks," *J. Wireless Net.*, vol. 8, pp. 169–185, 2002.

[272] A. Kashyap, K. Lee, M. Kalantari, S. Khuller, and M. Shayman, "Integrated topology control and routing in wireless optical mesh networks," *J. Comp. Netwks (Elsevier)*, vol. 51, no. 15, pp. 4237–4251, 2007.

[273] Y. Jahir, M. Atiquzzaman, H. Refai, and P. LoPresti, "Multipath hybrid ad hoc networks for avionics applications in disaster area," in *IEEE Conf. on Digital Avionics Sys. Conf.*, (Orlando, FL.), 2009.

[274] C. Intanagonwiwat, R. Govindan, and D. Estrin, "Directed diffusion: A scalable and robust communication paradigm for sensor networks," in *Proc AMC MobiCom*, (Boston, MA), pp. 56–67, 2000.

[275] D. Braginsky and D. Estrin, "Rumor routing algorithm for sensor networks," in *Int. Conf. on Distr. Comp. Sys.*, 2001.

[276] O. Awwad, A. Al-Fuqaha, B. Khan, and G. B. Brahim, "Topology control schema for better QoS in hybrid RF/FSO mesh networks," *IEEE Trans. Comm.*, vol. 60, pp. 1398–1406, 5 2012.

[277] R. Peach, G. Burdge, F. Reitberger, C. Visone, M. Oyler, C. Jensen, and J. Sonnenberg, "Performance of a 10 Gbps QoS-based buffer in a FSO/RF IP network," *Proc. SPIE, Free Space Laser Comm. X*, vol. 7814, 2010.

[278] B. Chen, K. Jamieson, H. Balakrishnan, and R. Morris, "SPAN: An energy-efficient coordination algorithm for topology maintenance in ad hoc wireless networks," *J. Wireless Net.*, vol. 8, no. 5, pp. 481–494, 2002.

[279] S. Sivathasan, "RF/FSO and LEACH wireless sensor networks: A case study comparing network performance," in *Proc. IEEE, Wireless and Opt. Comm. Net. (WOCN)*, pp. 1–4, 2009.

[280] S. Roychowdhury and P. Chiranjib, "Geographic adaptive fidelity and geographic energy aware routing in ad hoc routing," *Int J. Comput. & Comm. Tech.*, vol. 1, no. 2, pp. 309–313, 2010.

[281] Y. Yu, D. Estrin, and R. Govindan, "Geographical and energy-aware routing: A recursive data dissemination protocol for wireless sensor networks," Tech. Report: UCLA Computer Science Department-UCLA-CSD TR-01-0023, 2001.

[282] C. Intanagonwiwat, R. Govindan, D. Estrin, J. Heidemann, and F. Silva, "Directed diffusion for wireless sensor networking," *IEEE Trans. Net.*, vol. 11, no. 1, pp. 2–16, 2003.

[283] H. Shokrzadeh, A. T. Haghighat, F. Tashtarian, and A. Nayebi, "Directional rumor routing in wireless sensor networks," *Proc IEEE, Int. Conf. in Centr. Asia on Internet*, pp. 1–5, 2007.

[284] J. Faruque, K. Psounis, and A. Helmy, *Analysis of Gradient-Based Routing Protocols in Sensor Networks*, vol. 3560. Distributed Computing in Sensor Systems, 2005.

[285] A. Manjeshwar and D. P. Agrawal, "APTEEN: A hybrid protocol for efficient routing and comprehensive information retrieval in wireless sensor networks," *Proc IEEE, Int. Symp. Parall. n Distri Proc. (IPDPS)*, 2002.

[286] D. B. Johnson, D. A. Maltz, and J. Broch, "DSR: The dynamic source routing protocol for multi-hop wireless ad hoc networks," in *Ad Hoc Networking, edited by Charles E. Perkins, Chapter 5*, pp. 139–172, Addison-Wesley, 2001.

[287] Y. Bae Ko and N. H. Vaidya, "Location-aided routing (LAR) in mobile ad hoc networks," *J. Wireless Net. (Springer-Verlag)*, vol. 6, no. 4, pp. 307–321, 2000.

[288] R. Draves, J. Padhye, and B. Zill, "Routing in multi-radio, multi-hop wireless mesh networks," in *Proc. ACM, MobiCom*, 2004.

[289] V. D. Park and M. S. Corson, "A highly adaptive distributed routing algorithm for mobile wireless networks," in *Proc. IEEE, Comp. and Comm. Soc.(INFOCOM)*, 1997.

[290] M. R. Pearlman and Z. J. Hass, "Determining the optimal configuration for the zone routing protocol," *IEEE J. Sel. Areas in Comm.*, vol. 17, no. 8, pp. 1395–1414, 1999.

[291] C. E. Perkins, E. Belding-Royer, and S. Das, "Ad hoc on-demand distance vector (AODV) routing," Report, 2003.

[292] P. Mark and H. Wendi, "Sensor management policies to provide application QoS," *Proc. Elsevier, Ad Hoc Netwrks.*, vol. 1, pp. 235–246, 2003.

[293] E. Felemban, C. G. Lee, and E. Ekic, "MMSPEED: Multipath multispeed protocol for QoS guarantee of reliability and timelines in wireless sensor network," *IEEE Trans. Mob Comp.*, vol. 5, no. 6, pp. 738–754, 2006.

[294] S. Nesargi and R. Prakas, "A tunneling approach to routing with unidirectional links in mobile ad-hoc networks," in *Int. Conf. on Comp., Comm. and Netwrks*, 2000.

[295] S. A. Kah and S. A. Arshad, "QoS provisioning using hybrid FSO-RF based heirarchical model for wireless multimedia sensor networks," *J. Comp Sc. and Inf. Security*, vol. 4, no. 1, 2009.

[296] R. Shah and J. Rabaey, "Energy aware routing for low energy ad hoc sensor network," in *Proc IEEE WCNC 02, Orlando, FL*, pp. 350–355, 2002.

[297] K. Sohrabi, J. Gao, V. Ailawadhi, and G. J. Potie, "Protocols for self organization of a wireless sensor network," *IEEE Persn. Comm*, pp. 16–27, 2000.

[298] K. Akkaya and M. Younis, "An energy aware QoS routing protocol for wireless sensor networks," in *Int. Conf. on Distr. Comp. Sys. Workshop*, pp. 710–715, 2003.

[299] H. Tian, J. A. Stankovic, L. Chenyang, and T. Abdelzaher, "SPEED: A stateless protocol for real-time communication in sensor networks," in *Proc. IEEE, Int. Conf. Distr. Comp. Sys.*, pp. 46–55, 2003.

[300] X. Huang and Y. Fang, "Multiconstrained QoS multipath routing in wireless sensor networks," *J. Wireless Net.*, vol. 14, no. 4, pp. 465–478, 2008.

[301] N. Saxena, A. Roy, and J. Shin, "QUESt: A QoS-based energy efficient sensor routing protocol," *J. Wireless Comm. and Mobile Comp.*, vol. 9, no. 3, pp. 417–426, 2009.

[302] A. Majumdar, *Advanced Free Space Optics (FSO): A Systems Approach.* Springer, 2015.

[303] R. A. Nichols, A. R. Hammons, D. J. Tebben, and A. Dwivedi, "Delay tolerant networking for free-space optical communication systems," in *Proc IEEE, Sarnoff Symp.*, (Nassau Inn, Princeton, NJ), 2007.

[304] S. Burleigh, A. Hooke, L. Torgerson, K. Fall, V. Cerf, R. Durst, K. Scott, and H. Weiss, "Delay-tolerant networking: An approach to interplanetary internet," *IEEE Comm. Mag*, vol. 41, no. 6, pp. 128–126, 2003.

[305] L. Allen, M. W. Beijersbergen, R. J. C. Spreeuw, and J. P. Woerdman, "Orbital angular momemtum of light and the transformation of Laguerre-Gaussian laser modes," *Phys. Rev. A*, vol. 50, pp. 8185–8189, 1992.

[306] G. Gibson, J. Courtial, M. Padgett, M. Vasnetsov, V. Pas'ko, S. Barnett, and S. Franke-Arnold, "Free-space information transfer using light beams carrying orbital angular momentum," *Opt. Exp.*, vol. 12, pp. 5448–5456, 2004.

[307] J. Leach, J. Courtial, K. Skeldon, S. M. Barnett, S. Franke-Arnold, and M. J. Padgett, "Interferometric methods to measure orbital and spin, or the total angular momentum of a single photon," *Phys. Rev. Lett.*, vol. 92, no. 1, pp. 013601–1–013601–4, 2004.

[308] M. T. Gruneosen, W. A. Miller, R. C. Dymale, and A. M. Seiti, "Holographic generation of complex fields with spatial light modulators: Application to quantum key distribution," *Appl. Opt.*, vol. 47, pp. A32–A42, 2008.

[309] J. Wang, J. Y. Yang, I. M. Fazal, N. Ahmed, Y. Yan, H. Huang, Y. Ren, Y. Yue, S. Dolinar, M. Tur, and A. E. Willner, "Terabit free-space data transmission employing orbital angular momentum multiplexing," *Nat. Photon.*, vol. 6, no. 7, pp. 488–496, 2012.

[310] H. Huang, G. Xie, Y. Yan, N. Ahmed, Y. Ren, Y. Yue, D. Rogawski, M. Tur, B. Erkmen, K. Birnbaum, S. Dolinar, M. Lavery, M. J. Padgett, and A. E. Willner, "100 Tbit/s free-space data link using orbital angular momentum mode division multiplexing combined with wavelength division multiplexing," in *OFC/NFOEC Tech. Digest*, 2013.

[311] J. A. Anguita, M. A. Neifeld, and B. V. Vasic, "Turbulence-induced channel crosstalk in an orbital angular momentum multiplexed free-space optical link," *Appl. Opt.*, vol. 47, no. 13, pp. 2414–2429, 2008.

[312] C. Paterson, "Atmospheric turbulence and orbital angular momentum of single photons for optical communication," *Phys. Rev. Lett.*, vol. 94, no. 15, p. 153901, 2005.

[313] Y. Ren, H. Huang, G. Xie, N. Ahmed, Y. Yan, B. I. Erkmen, and et al., "Atmosphetic turbulence effects on the performance of a free space optical link employing orbital angular momemtum multiplexing," *Opt. Lett.*, vol. 38, no. 20, pp. 4062–4065, 2013.

[314] Z. J. Zhao, R. Liao, S. D. Lyke, and M. C. Roggemann, "Reed-Solomon coding for free-space optical communication through turbulent atmosphere," *Proc IEEE, Aerospace Conf.*, pp. 1–12, 2010.

[315] I. B. Djordjevic and M. Arabaci, "LDPC-coded orbital angular momentum (OAM) modulation for free-space optical communication," *Opt. Exp.*, vol. 18, no. 24, pp. 24722–24728, 2010.

[316] S. M. Zhao, B. Wang, L. Y. Gong, Y. B. Sheng, W. W. Cheng, X. L. Dong, and B. Y. Zheng, "Improving the atmosphere turbulence tolerance in holographic ghost imaging system by channel coding," *J. Lightwave Tech.*, vol. 31, no. 17, pp. 2823–2828, 2013.

[317] Y. Ren, H. Huang, G. Xie, and et al., "Simultaneous pre-and post-turbulence compensation of multiple orbital-angular-momentum 100-Gbit/s data channels in a bidirectional link using a single adaptive-optics system," *Frontiers in Optics (OSA)*, 2013.

[318] S. M. Zhao, J. Leach, L. Y. Gong, J. Ding, and B. Y. Zheng, "Aberration corrections for free-space optical communications in atmosphere turbulence using orbital angular momentum states," *Opt. Exp.*, vol. 20, pp. 452–461, 2012.

[319] S. Zhao, B. Wang, L. Zhou, L. Gong, W. Cheng, Y. Sheng, and B. Zheng, "Turbulence mitigation scheme for optical communications using orbital angular momentum multiplexing based on channel coding and wavefront correction." Cornell University Library: Optics, Jan. 2014.

[320] K. Hirabayashi, T. Yamamoto, and S. Hino, "Optical backplane with free-space optical interconnections using tunable beam deflectors and a mirror for bookshelf-assembled terabit per second class asynchronous transfer mode switch," *Opt. Eng.*, vol. 37, pp. 1332–1342, 2004.